\documentclass[10pt]{JHEP3} 

\usepackage{epsfig,multicol,bbm}

\newcommand\fverb{\setbox\fverbbox=\hbox\bgroup\verb}
\newcommand\fverbdo{\egroup\medskip\noindent%
            \fbox{\unhbox\fverbbox}\ }
\newcommand\fverbit{\egroup\item[\fbox{\unhbox\fverbbox}]}
\newbox\fverbbox

\usepackage{amsmath}
\usepackage{amssymb}
\usepackage{amsfonts}
\usepackage{latexsym}
\usepackage{graphicx}
\usepackage{dcolumn}
\usepackage{bm}
\usepackage{empheq}
\usepackage{times} 
\newcommand{\eq}[1]{\begin{equation}#1\end{equation}}

\newcommand{\ea}[1]{\begin{equation}\begin{aligned}#1\end{aligned}\end{equation}}
\newcommand{\itm}[1]{\begin{itemize}#1\end{itemize}}
\newcommand{\od}[2]{\frac{\textrm{d} #1}{\textrm{d} #2}}  
\newcommand{\pd}[2]{\frac{\partial #1}{\partial #2}}  
\newcommand{\lrp}[1]{\left( #1 \right)}  
\newcommand{\lrb}[1]{\left( #1 \right)}  
\newcommand{\lrsb}[1]{\left[ #1 \right]}  
\newcommand{\lrab}[1]{\left\langle #1 \right\rangle}  
\newcommand{\lrmb}[1]{\left| #1 \right| } 
\def\la{\langle}
\def\ra{\rangle}

\def\rd{\partial}
\def\vk{\bm{k}}

\def\dop{\dot{\phi}}

\def\qsg{Q_{\sigma}}
\def\qs{Q_{s}}

\def\ce{c_{\textrm{e}}}
\def\D{\mathcal{D}}
\newcommand{\ab}[1]{\langle #1 \rangle} 
\def\I{\textrm{I}}
\newcommand{\soev}[3]{\lrab{ #1 \lrmb{ #2}#3 }} 

\title{On Cross-correlations between Curvature and Isocurvature Perturbations during Inflation}

\author{Xian Gao\\
    Key Laboratory of Frontiers in Theoretical Physics,\\
    Kavli Institute for Theoretical Physics China, \\
    Chinese Academy of
    Sciences, Beijing 100190, P.R.China\\
    Email:\email{gaoxian@itp.ac.cn}}

\preprint{CAS-KITPC/ITP-148}

\keywords{Cosmological perturbation theory, Curvature perturbation,
Isocurvature perturbation, Non-Gaussianity}

\abstract{
   We investigate the effects of couplings between curvature and isocurvature
perturbations before and around horizon-crossings during
cosmological
 inflation. We consider a generalized two-field
inflation model, in which the non-canonical kinetic term allows us
arbitrary sound speeds
 of curvature and isocurvature perturbations. By using the field-theoretical perturbative analysis,
we calculate the cross-spectrum between curvature and isocurvature
perturbations and
 the corrections to curvature and isocurvature power spectra due to
 the presence of couplings between them. Our analysis confirms
 previous results that the cross-correlations are generated
 and amplified when perturbations cross the horizons.
Moreover, we find the cross-correlation, which was previously shown
to be first-order in slow-roll parameter, can be enhanced when the
sound speed of isocurvature perturbation is much smaller than that
of the curvature perturbation. This is because in this case the
isocurvature perturbation exits its horizon much earlier than the
curvature perturbation and acts as a nearly constant source on the
curvature perturbation. }

\begin{document}


\section{Introduction}

It is believed that the large-scale structure in our universe grows
up from the primordial quantum fluctuations during a period of
cosmological inflation (see e.g. \cite{Lyth:1998xn} for a review).
The predictions of inflation have been supported by current
observational data \cite{Komatsu:2008hk}.

The simplest model for inflation is based on the picture that a
single scalar field rolls down its potential, making the universe
inflate and also generating quantum fluctuations. However, various
alternatives are investigated extensively, one of which is
multi-field inflation models
\cite{Langlois:1999dw,Amendola:2001ni,Wands:2002bn,Kofman:1986wm,Mukhanov:1991rp,Polarski:1992dq,Wands:1996kb,GarciaBellido:1996ke,Mukhanov:1997fw,Langlois:1999dw,Starobinsky:2001xq,vanTent:2003mn,Gordon:2000hv,DiMarco:2002eb,DiMarco:2005nq,Bartolo:2001vw,Byrnes:2006fr,Tsujikawa:2002qx,Langlois:2006vv,Langlois:2005ii,Langlois:2005qp,Starobinsky:1994mh,Bassett:2005xm,Langlois:2008mn,Lalak:2007vi,RenauxPetel:2008gi}.
The goal of the studies of multi-field models is two-fold. Firstly,
many inflation models based on particle physics or string theory
usually involve many scalar fields, which can also have
non-canonical kinetic terms. Secondly, it has been clear that any
detection of primordial non-gaussianity would rule out the simplest
slow-roll single field inflation models{\footnote{Meanwhile, there
indeed exist various secondary effects which can also produce a
detectable level of non-Gaussianity in the observed Cosmic Microwave
Background and Large-scale Structure.}}. On the other hand, multiple
field models provide us more possibilities and have been discussed
extensively
\cite{Gao:2008dt,Xue:2008mk,Gao:2009gd,Gao:2009at,Huang:2009xa}.
Thus, it is natural and important to study multi-field inflation
models in details.

However, in the context of multi-field models, except for a few
specific models, even the predictions for the spectra of primordial
perturbations are a non-trivial task. The main reason is that, there
are couplings between adiabatic mode and entropy mode(s), even at
linear level. A well-known result is that the curvature (or
adiabatic) perturbation can evolve on super-horizon scales in
multi-field inflation whereas it is conserved in single-field
inflation. This is due to that the entropy (isocurvature)
perturbation modes act as a source term in the evolution equation
for the curvature perturbation. This phenomena was first emphasized
in \cite{Starobinsky:1994mh}. The production of adiabatic and
entropy modes for two-field models with a generic potential was
studied in \cite{Gordon:2000hv} where a decomposition into
instantaneous adiabatic/entropy modes was firstly introduced.
Multi-field models with non-canonical kinetic terms have been
investigated in the slow-rolling approximation in
\cite{Mukhanov:1997fw,Starobinsky:2001xq,Langlois:2008mn}, where the
adiabatic/entropy decomposition technique was also extended in
\cite{DiMarco:2002eb,DiMarco:2005nq} with non-canonical kinetic
terms.

As has been stressed above, one of the difficulties in multi-field
inflation researches is that, one cannot trace back the adiabatic
mode to one of these scalar fields, and the entropy modes to
remaining scalar fields. In general, all relevant scalar fields are
mixed together to give one adiabatic modes and $N-1$ entropic modes.
On the other hand, this is also the reason why we should expect
cross-correlations between adiabatic and entropy modes
\cite{Langlois:1999dw,Amendola:2001ni,Wands:2002bn}. However, in
most of the previous works, \emph{quantum cross-correlations}
between adiabatic and entropy modes before and around
horizon-crossing $\lrab{\qsg\qs}_{\ast}$ are expected to be small,
based on the observation that the cross-correlations are of order
slow-rolling parameters before and around horizon-crossing. This
cross-correlation has been studied analytically in
\cite{Bartolo:2001vw} as a phenomena of oscillations between two
perturbation modes and also been investigated in details in
\cite{Byrnes:2006fr} with canonical kinetic term and in
\cite{Lalak:2007vi} with non-canonical kinetic term. Numerical
studies were also presented in \cite{Tsujikawa:2002qx,Choi:2007su}.

The goal of this work is to study the cross-correlations in details.
The analytic treatments in
\cite{Bartolo:2001vw,Byrnes:2006fr,Lalak:2007vi} are based on the
``diagonalization" of the coupled system: time-dependent orthogonal
matrices are introduced to abstract the approximately decoupled
degrees of freedom which should be quantized. Indeed, this is the
standard treatment to a coupled system. In this note, we take a
slightly alternative approach --- that is, we treat the couplings
between adiabatic and entropy modes as ``two-point" interactions,
and use standard field theoretical perturbative methods to calculate
the cross-correlation and also the corrections to adiabatic/entropy
spectra themselves. More precisely, we split the full
quadratic-order action into a ``free" part in which adiabatic mode
and entropy mode decouple with each other and a ``two-point"
interaction part, where the two-point couplings are treated as
interaction vertices. Generally speaking, this approach supplies us
a systematic perturbative procedure to study the coulings between
adiabatic and entropy modes in details, especially in the cases
where the ``diagonalization" of the coupled system cannot be done
easily{\footnote{Especially, in models as considered in this paper
with different adiabatic and entropy speeds of sound, i.e. $c_a \neq
c_e$, the diagonalization is a non-trivial task. Since in this case,
the system is equivalent to a coupled oscillator system with
different (free-theory) energy eigenvalues ($c_ak\neq c_e k$) plus
``time-dependent" interactions. The simultaneous digonalization of
both the free-theory Hamiltonian and the time-dependent interaction
is not trivial, and in this case the traditional effective method is
the perturbation theory. Models considered in
\cite{Bartolo:2001vw,Byrnes:2006fr,Lalak:2007vi} has the same $c_a$
and $c_e$, in the words of quantum oscillators, the couple two-state
system has degenerate (free) energy eigenstates, in which the
diagonalization can be done easily.}} or the ``time-independence" of
the diagonolization matrices is not a good approximation.

We consider a generalized two-field inflation model, as described in
the next section. The prototype of this form of Lagrangian was
proposed in \cite{Langlois:2008wt,Arroja:2008yy} and includes
multi-field $k$-inflation \cite{Langlois:2008mn,Gao:2008dt} and
two-field DBI model
\cite{Langlois:2008wt,Langlois:2008qf,Langlois:2009ej,Gao:2009gd,Mizuno:2009cv,RenauxPetel:2009sj,Mizuno:2009mv}
as special case and thus deserves detailed study. Actually,
non-gaussianities in multi-field models with non-canonical kinetic
terms have been extensively investigated in
\cite{Langlois:2008wt,Arroja:2008yy,Gao:2008dt,Gao:2009gd,Gao:2009at,Langlois:2008qf,Langlois:2009ej,Kawakami:2009iu,Mizuno:2009cv,Lehners:2009ja,Byrnes:2009qy,RenauxPetel:2009sj,Mizuno:2009mv}
(see also
\cite{Tye:2008ef,Ji:2009yw,Li:2008jn,Xue:2008mk,Li:2008qv,Huang:2009vk,Huang:2009xa}).
Although the main task in this note is not to study a complex
multi-field model, this generalized Lagrangian can make our analysis
of the cross-correlation in a more general background. Especially,
the non-canonical kinetic term of our model allows us arbitrary
sound speeds of adiabatic and entropy perturbations, which are
essential for our following analysis. Our work can be viewed as
generalization of the analysis in
\cite{Bartolo:2001vw,Byrnes:2006fr,Lalak:2007vi} to a general class
of two-field models with non-canonical kinetic terms and arbitrary
speeds of sound for adiabatic and entropy modes, $c_a \neq c_e$.

This paper is organized as follows. In the next section we introduce
a generalized two-field inflation model, and describe the scalar
perturbations of it. The third section is devoted to investigate the
cross-correlations in detail, based on the field-theoretical
perturbative approach. The last section is devoted to conclusion and
discussion on the limitation and possible extension of this work.

\section{Generalized Two-field Inflation Model}

In this work, we consider a very general class of two-field
inflation models with action of the form:
    \eq{{\label{model}}
        S = \int d^4x \sqrt{-g} \lrsb{ \frac{1}{2}M_{\textrm{p}}^2 R - P(X,Y,\phi^I)
        }  \,,
    }
where $X\equiv X^I_I = G_{IJ}X^{IJ}$ and $Y\equiv X^I_J X^J_I $ with
$X^{IJ} \equiv -\frac{1}{2}g^{\mu\nu} \partial_{\mu}\phi^I
\partial_{\nu}\phi^J$, $G_{IJ}$ is the metric for the field space and $M_{\textrm{p}} \equiv 1/\sqrt{8\pi G}$ is the reduced Planck mass which we set to unity in the following.
 In two-field
case, all higher order contractions among $X_{IJ}$'s can be
expressed in terms of $X$ and $Y$, e.g.,
    \[
    \begin{aligned}
        X^I_J X^J_K X^K_I &= -\frac{X^{3}}{2}+\frac{3}{2}XY \,,\\
        X^I_J X^J_K X^K_L X^L_I &= -\frac{1}{2}X^{4}+X^{2}Y+\frac{1}{2}Y^{2}
        \,,
    \end{aligned}
    \]
etc. The model (\ref{model}) includes multi-field $k$-inflation and
two-field DBI model as special cases. For example, in multi-DBI
model the Lagrangian is $P = - \frac{1}{f(\phi^I)} \lrp{ \sqrt{
\mathcal{D} }-1} -
        V(\phi^I)$
with
    \ea{
        \mathcal{D} &\equiv \det \lrp{ G^I_J - 2f X^I_J } \\
        &= 1 -2f G_{IJ}X^{IJ} + 4f^2 X^{[I}_I X^{J]}_J - 8f^3
        X^{[I}_I X^J_J X^{K]}_K + 16 f^4 X^{[I}_I X^J_J X^K_K
        X^{L]}_L \,.
    }
This expression for determinant $\mathcal{D}$ is general. In this
work, we focus on two-field case, thus the last two terms exactly
vanish, leaving us effectively $\mathcal{D} \equiv 1 -2f
G_{IJ}X^{IJ} + 4f^2 X^{[I}_I
        X^{J]}_J $. In terms of (\ref{model}), this is just $\mathcal{D} = 1-2fX+2f^2\lrp{X^2-Y}$.

This form of scalar-field Lagrangian in (\ref{model}) is the most
general Lagrangian for two-field models and thus deserves detailed
investigations. The goal of choosing such a general Lagrangian in
this note is not only because recent investigations on
non-Gaussianities in multi-field are based on some similar
Lagrangian
\cite{Langlois:2008wt,Arroja:2008yy,Gao:2008dt,Gao:2009gd,Gao:2009at,Langlois:2008qf,Langlois:2009ej,Kawakami:2009iu,Mizuno:2009cv,Lehners:2009ja,Byrnes:2009qy,RenauxPetel:2009sj,Mizuno:2009mv},
but also in order to see the effects on perturbations from the
structure of the theory in a wider range{\footnote{Actually the
Lagrangian in (\ref{model}) was motivated from some similar models
in previous investigations. For example, in
\cite{Langlois:2008mn,Gao:2008dt} a Lagrangian of the form
$P(X,\phi^I)$ was introduced, which described a multi-field
generalization of single-field $k$-inflation. In
\cite{Arroja:2008yy} a special form $ \tilde{P}(\tilde{Y},\phi^I) $
with $\tilde{Y} \equiv X + \frac{b(\phi^I)}{2} \lrp{ X^2 -
X_{IJ}X^{IJ} }$ was chosen in the investigation of bispectrua in
two-field models.}. As we will see, the non-canonical kinetic term
supplies us two different speeds of sound for adiabatic and entropy
modes which we denote as $c_a$ and $c_e$ respectively, which are
essential for our following analysis.

\subsection{Background Equations of Motion}

In this work, we investigate scalar perturbations around a flat FRW
background, the background spacetime metric takes the form
    \eq{{\label{metric}}
        ds^2 = -dt^2 + a^2(t) \delta_{ij} dx^i dx^j \,,
    }
where $a(t)$ is the  scale-factor. The Friedmann equation and the
continuity equation are
    \ea{
        H^2 &\equiv  \frac{\rho}{3} = \frac{1}{3} \lrp{ 2X^{IJ} P_{,\ab{IJ} } - P
        } \,,\\
        \dot{\rho} &= -3H( \rho + P ) \,,
    }
where and in what follows we denote $P_{,\ab{IJ}} \equiv
\pd{P}{X^{IJ}}$, $P_{,\ab{IJ}\ab{KL}} \equiv \frac{\partial^2
P}{\partial X^{IJ} \partial X^{KL}}$ etc. for short. In the above
equations, all quantities are evaluated on the background. From the
above two equations we can also get another convenient equation
    \eq{{\label{eq_dotH}}
        \dot{H} = -X^{IJ} P_{,\ab{IJ}} \,.
    }
The background equations of motion for the scalar fields are
    \eq{
        P_{,\ab{IJ}} \ddot{\phi}^I + \lrp{ 3HP_{,\ab{IJ}} + \dot{P}_{,\ab{IJ}}
        } \dop^I - P_{,J} = 0\,,
    }
where $P_{,I}$ denotes derivative of $P$ with respect to $\phi^I$:
$P_{,I} \equiv \pd{P}{\phi^I}$.

In this work, we investigate cosmological perturbations during an
exponential inflation period. Thus, from (\ref{eq_dotH}) it is
convenient to define a slow-roll parameter for the expansion rate
    \eq{{\label{def_epsilon}}
        \epsilon \equiv -\frac{d\ln H}{d \ln a}= -\frac{\dot{H}}{H^2} = \frac{P_{,\ab{IJ}} \dop^I
        \dop^J}{2H^2} \,.
    }
In this note we do not go into details of solving the background
equations of motion, but only assume that the structure of
$P(X,Y,\phi^I)$ and thus the background dynamical equations permit
such an exponential expansion period.

\subsection{Linear Perturbations}

In this work we focus on the linear perturbations. In multi-field
models, it is convenient to work in spatially-flat gauge, where the
metric (scalar sector) is unperturbed as in (\ref{metric}), and the
perturbation of the system is encoded in the perturbations of the
scalar fields, which we denote $\delta\phi^I \equiv Q^I$ for short.

In multi-field inflation models, it is convenient to decompose
perturbations into instantaneous adiabatic and entropy perturbations
\cite{Gordon:2000hv,Bassett:2005xm}. This decomposition was firstly
introduced in \cite{Gordon:2000hv} in the study of two-field
inflation with a generic potential, and was extended in
\cite{Starobinsky:2001xq,DiMarco:2002eb,DiMarco:2005nq} in two-field
models with non-canonical kinetic terms. This decomposition
technique was also generalized to non-linear perturbations
\cite{Langlois:2006vv} in the context of covariant non-linear
formalism \cite{Langlois:2005ii,Langlois:2005qp}.

The ``adiabatic direction" corresponds to the direction of the
``background inflaton velocity", for model described in this work,
it is
    \eq{
        e^I_{\sigma} \equiv \frac{\dop^I}{ \sqrt{ P_{,\ab{JK}} \dop^J\dop^K
        }} \equiv \frac{\dop^I}{\dot{\sigma}}\,,
    }
where $\dot{\sigma}$ is defined as
    \eq{{\label{dot_sigma}}
        \dot{\sigma} \equiv \sqrt{ P_{,\ab{JK}} \dop^J\dop^K}  \,,
    }
which is the generalization of the background inflaton velocity.
Actually $\dot{\sigma}$ is essentially a short notation and has
nothing to do with any concrete field. Note that $\dot{\sigma}$ is
related to the slow-roll parameter $\epsilon$ as $\dot{\sigma}^2 =
2H^2 \epsilon$.

In this work we focus on two-field case. We introduce the entropy
basis $e^I_s$ which is orthogonal to $e^I_{\sigma}$. The orthogonal
condition can be defined as{\footnote{In specified models, other
choices of orthogonal conditions are possible. The idea is to make
the kinetic terms of the perturbations decouple. The final results
for curvature/isocurvature perturbations is independent of different
choices of orthogonal conditions.}}
    \eq{{\label{orthogonal}}
        P_{,\ab{IJ}} e^I_m e^J_n \equiv \delta_{mn} \,,\qquad\qquad
        m,n=\sigma,s
    }
Thus the scalar-field perturbation $Q^I$ can be decomposed into
instantaneous adiabatic/entropy modes as:
    \eq{
        Q^I \equiv e^I_{\sigma} Q_{\sigma} + e^I_{s} Q_{s}  \,.
    }

In spatially-flat gauge, the quadratic-order action for the
perturbations of the model (\ref{model}) can be calculated
straightforwardly. After instantaneous adiabatic/entropy modes
decomposition, up to total derivative terms, the second-order action
for the scalar perturbations takes the form
\cite{Langlois:2008wt,Langlois:2008qf,Arroja:2008yy,Gao:2009at}
 \eq{{\label{S2_Q}}
    S_{2}\equiv\int dtd^{3}xa^{3}\left[\frac{1}{2}K_{mn}\dot{Q}^{m}\dot{Q}^{n}-\delta_{mn}\frac{1}{2a^{2}}\partial_{i}Q^{m}\partial_{i}Q^{n}+\Xi_{mn}\dot{Q}^{m}Q^{n}-\frac{1}{2}M_{mn}Q^{m}Q^{n}\right] \,,\\
} with \ea{{\label{K_mn}}
    K_{mn} &\equiv \delta_{mn} + \dot{\sigma}^2 P_{,\ab{IK} \ab{JL}} e^I_{\sigma} e^K_n e^J_{\sigma} e^L_m \,,\\
        &= \delta_{mn} + \lrp{ \frac{1}{c_a^2}-1 }
        \delta_{\sigma m} \delta_{\sigma n} + \lrp{ \frac{1}{c_e^2} - 1 }
        \lrp{\delta_{mn}- \delta_{\sigma m} \delta_{\sigma n}
        } \,,\\
    \Xi_{mn} &\equiv \mathcal{N}_{IJ}e_{m}^{I}e_{n}^{J}+\left(P_{\left\langle IJ\right\rangle }+2P_{\left\langle IK\right\rangle \left\langle JL\right\rangle }X^{KL}\right)e_{m}^{I}\dot{e}_{n}^{J} \,,\\
    -M_{mn} &= -\mathcal{M}_{IJ}e_{m}^{I}e_{n}^{J}+2\mathcal{N}_{IJ}\dot{e}_{m}^{I}e_{n}^{J}+\left(P_{\left\langle IJ\right\rangle }+2P_{\left\langle IK\right\rangle \left\langle JL\right\rangle
    }X^{KL}\right)\dot{e}_{m}^{I}\dot{e}_{n}^{J} \,, \qquad\quad
    (m,n=\sigma,s)
        }
where
    \ea{
        \mathcal{N}_{IJ} &\equiv \dot{\phi}^{K}P_{\left\langle KI\right\rangle ,J}-\frac{2}{H}X^{KL}X^{NM}P_{\left\langle NI\right\rangle \left\langle KL\right\rangle }P_{\left\langle MJ\right\rangle } \,,\\
        -\mathcal{M}_{IJ} &\equiv P_{,IJ}+\left(X^{MN}P_{\left\langle MN\right\rangle }+2X^{MN}X^{PQ}P_{\left\langle MN\right\rangle \left\langle PQ\right\rangle }-3H^{2}\right)\frac{X^{KL}}{H^{2}}P_{\left\langle KI\right\rangle }P_{\left\langle LJ\right\rangle }\\
        &\qquad +\frac{1}{H}\left(P_{,I}-2X^{MN}P_{\left\langle MN\right\rangle ,I}\right)P_{\left\langle KJ\right\rangle }\dot{\phi}^{K}+\frac{1}{a^{3}}\frac{d}{dt}\left(\frac{a^{3}}{H}X^{KL}P_{\left\langle KI\right\rangle }P_{\left\langle LJ\right\rangle
        }\right) \,.
    }
In (\ref{K_mn}) we introduce
    \ea{{\label{def_ca_ce}}
        c_a^2 &\equiv \frac{P_{,X}+2XP_{,Y}}{P_{,X}+2X\left(P_{,XX}+4XP_{,XY}+3P_{,Y}+4X^{2}P_{,YY}\right)} \,,\\
        c_e^2 &\equiv \frac{P_{,X}}{P_{,X}+2XP_{,Y}} \,,
    }
which are the propagation speeds of the adiabatic mode and entropy
mode respectively. From (\ref{K_mn}), the kinetic term $K_{mn}
\dot{Q}^m \dot{Q}^n$ has been diagonalized, as a result of
adiabatic/entropy decomposition.

For our purpose in this work, the different speeds of sound for
adiabatic and entropy modes are essential for the investigation of
cross-correlations. Actually in multi-field models, it is generic
fact that $c_a \neq c_e$ which was firstly point out apparently in
\cite{Easson:2007dh,Huang:2007hh} in the investigation of brane
inflation model. In multiple $k$-inflation with Lagrangian of the
form $P(X,\phi^I)$, the adiabatic mode propagate with sound speed
$c_s$ while entropy modes propagate with the speed of light
\cite{Langlois:2008mn,Gao:2008dt}. While subsequently in
\cite{Langlois:2008wt,Arroja:2008yy}, it was shown that in multi-DBI
models, the adiabatic mode and entropy modes propagate with the same
speed of sound. Now it becomes clear that, in multi-field
inflationary models, adiabatic mode and entropic modes in general
propagate with different speeds of sound $c_a$ and $c_e$, which
depend on the structure of specific theory
\cite{Langlois:2008wt,Arroja:2008yy} (see also
\cite{Arroja:2008yy,Langlois:2008qf,Cai:2008if,Cai:2009hw,Ji:2009yw,Gao:2008dt,Gao:2009gd}
for extensive investigations on general multi-field models with
different $c_a$ and $c_e$).

It is now convenient to introduce the canonically normalized
variables
    \eq{
        \tilde{Q}_{\sigma} = \frac{a}{c_a}\qsg\,,\qquad \tilde{Q}_{s} =
        \frac{a}{c_e} \qs\,.
    }
After straightforward but tedious calculations, the quadratic action
for $\tilde{Q}_{\sigma}$ and $\tilde{Q}_s$ (after using conformal
time $\eta$ defined by $dt=ad\eta$ and up to total derivative terms)
takes the form:
    \ea{{\label{action_u}}
        S_2[\tilde{Q}_{\sigma}, \tilde{Q}_s ] &= \int d\eta d^3x\, \frac{1}{2} \left[ \tilde{Q}_{\sigma}'^2 - c_a^2 \lrb{\rd \tilde{Q}_{\sigma}}^2 +  \frac{z''}{z}  \tilde{Q}_{\sigma}^2 + \tilde{Q}_s'^2 - c_e^2 \lrb{\rd \tilde{Q}_s}^2 + \lrb{ \frac{\alpha''}{\alpha} - a^2\mu_s^2 } \tilde{Q}_s^2 \right. \\
        &\qquad\qquad\qquad\qquad \left. -2 \mathcal{H}\xi \, \tilde{Q}_{\sigma}' \tilde{Q}_s + 2 \frac{z' \mathcal{H}\xi }{z}\, \tilde{Q}_{\sigma} \tilde{Q}_s
        \right] \,,
    }
where
    \ea{{\label{def_para}}
        \xi & = \frac{1}{\sqrt{2\epsilon} \sqrt{P_1} c_a} \lrsb{ (1 + c_a^2) \frac{P_{,s}}{H^2} - 2 \epsilon c_a^2
        P_{,1s}
    } \,,\\
        z &=  a \frac{\dot{\sigma}}{c_a H} \,,\\
        \alpha &= a \sqrt{P_1} \,,\\
        \mu_s^2 &=  -\frac{P_{,ss}}{P_1} + \frac{\dot{\sigma}^2\tilde{R}}{2P_1}
         - \frac{P_{,s}^2 }{c_a^2 \dot{\sigma}^2 P_1^2}  + \frac{2 P_{1s} P_{,s} }{P_1^2} \,,
    }
with
    \ea{
        P_{,s} &= P_{,I}e^I_s \sqrt{ P_1
        }\,
        c_e\,,\qquad P_{,1s} = (\partial_I P_1) e^I_s  \sqrt{
        P_1
        }\,
        \ce\,,\qquad P_{,ss} = c_e^2 P_1  \lrp{\D_I\D_J P} e^I_s e^J_s
        \,,\\
        P_{1} &\equiv  P_{,X}+2XP_{,Y} \,,
    }
and $\tilde{R}$ is the Ricci scalar of field space metric $G_{IJ}$,
$\mathcal{D}_I$ is the covariant derivative associated to $G_{IJ}$
(i.e. $\mathcal{D}_I\mathcal{D}_J P \equiv P_{,IJ} - \Gamma^K_{IJ}
P_{,K}$). Note that these various parameters are evaluated on the
background. In general, the time-dependence of these various
parameters are complicated. In this note, in order to proceed, we
introduce several slow-varying parameters:
    \ea{{\label{sv_para}}
        \eta_{\epsilon} \equiv \od{\ln\epsilon}{\ln a}\,,\qquad s_a
        \equiv \od{\ln c_a}{\ln a} \,,\qquad s_e
        \equiv \od{\ln c_e}{\ln a}\,,\qquad \eta_p
        \equiv \od{\ln P_1}{\ln a}\,, \qquad \eta_{\xi}
        \equiv \od{\ln \xi}{\ln a}
    }
where $a$ is the scale-factor.

In canonical quantization procedure, the quantum fields are
decomposed as
    \eq{{\label{mode_expansion}}
        \tilde{Q}_{\sigma}(\vk,\eta) \equiv a_{\vk} u_{\sigma}(k,\eta) + a^{\dag}_{-\vk}
        u^{\ast}_{\sigma}(k,\eta) \,, \qquad\qquad \tilde{Q}_s(\vk,\eta) \equiv b_{\vk} u_s(k,\eta) + b^{\dag}_{-\vk}
        u^{\ast}_s(k,\eta) \,,
    }
The equations of motion for the mode functions $u_{\sigma}$ and
$u_s$ can be get from varying (\ref{action_u}):
    \ea{{\label{eom_pert_original}}
        u''_{\sigma}  +  \lrp{c_a^2 k^2 - \frac{z''}{z}}
        u_{\sigma} - \mathcal{H}\xi u'_s - \frac{(z\mathcal{H}\xi)'}{z} u_s &=0 \,,\\
        u''_s  + \lrp{ c_e^2 k^2 - \frac{\alpha''}{\alpha} + a^2\mu_s^2
        } u_s + \mathcal{H}\xi u'_{\sigma} - \frac{z'\mathcal{H}\xi}{z} u_{\sigma} &=0 \,,
    }
These two equations form a closed system for the scalar
perturbations.

\section{Perturbative Analysis}

As described in the Introduction, the idea in this paper is to treat
the coupling between adiabatic and entropy modes as ``two-point"
interaction vertices, and to use field theoretical perturbative
approaches to evaluate the cross-correlations.

\subsection{Interaction Hamiltonian}

The first line in (\ref{action_u}) describes a decoupled two-field
system, where the two decoupled modes can be quantized
independently. While the second-line can be identified as the
``two-point cross-interaction vertices":
    \eq{
         S_c[\tilde{Q}_{\sigma}, \tilde{Q}_s] = \int d\eta d^3x\, \lrb{ -\mathcal{H}\xi\, \tilde{Q}_{\sigma}' \tilde{Q}_s +  \frac{z' \mathcal{H}\xi}{z}\, \tilde{Q}_{\sigma}
         \tilde{Q}_s } \,,
    }
where the dimensionless cross-coupling $\xi$ is given in
(\ref{def_para}). In the operator formalism of quantization,
interaction Hamiltonian is needed. The Hamiltonian density which is
defined by $\mathcal{H}\equiv \pi_a Q'_a - \mathcal{L}$ can be split
into two parts: $\mathcal{H} \equiv\mathcal{H}_0 + \mathcal{H}_c $,
with
    \ea{{\label{Hamiltonian}}
        \mathcal{H}_0 &\equiv \frac{1}{2}\pi_{\sigma}^2 + \frac{1}{2} c_a^2 \lrb{\rd
        \tilde{Q}_{\sigma}}^2 -\frac{z''}{2z} \tilde{Q}_{\sigma}^2 +
        \frac{1}{2}\pi_s^2 + \frac{1}{2} c_e^2 \lrb{\rd
        \tilde{Q}_s}^2  -
        \frac{1}{2} \lrb{ \frac{\alpha''}{\alpha} - a^2 \mu_s^2
        -\mathcal{H}^2\xi^2
        } \tilde{Q}_s^2 \,,\\
        \mathcal{H}_{c} &\equiv \mathcal{H}\xi\, \pi_{\sigma} \tilde{Q}_s -
        \frac{z'\mathcal{H}\xi}{z} \tilde{Q}_{\sigma} \tilde{Q}_s \,,
    }
where $\mathcal{H}_0$ describes decoupled system while
$\mathcal{H}_c$ describes the cross interactions. From
$\mathcal{H}_0$, the free-theory canonical momenta are (in
interaction picture) are related with time-derivatives of the fields
as
    \ea{
        \tilde{Q}'_{\sigma} \equiv \pd{\mathcal{H}_0}{\pi_{\sigma}}
        = \pi_{\sigma}\,,\qquad \tilde{Q}'_{s} \equiv \pd{\mathcal{H}_0}{\pi_{s}}
        = \pi_{s} \,,
    }
thus in the interaction picture, the cross-interaction vertices can
be written in terms of $\tilde{Q}_m$ and $\tilde{Q}'_m$ as
    \eq{{\label{Hc}}
        \mathcal{H}_c = \mathcal{H}\xi\, \tilde{Q}'_{\sigma} \tilde{Q}_s -
        \frac{z'\mathcal{H}\xi}{z} \tilde{Q}_{\sigma} \tilde{Q}_s \,.
    }

Since $\tilde{Q}_{\sigma}$ and $\tilde{Q}_s$ are the canonical
variables for quantization, the corresponding mode functions in
(\ref{mode_expansion}) satisfy the decoupled (``free-theory")
equations of motion:
    \ea{{\label{eom_pert_dec}}
        u''_{\sigma}  +  \lrp{c_a^2 k^2 - \frac{z''}{z}}
        u_{\sigma}  &=0 \,,\\
        u''_s  + \lrp{ c_e^2 k^2 - \frac{\alpha''}{\alpha} + a^2\mu_s^2
        } u_s  &=0 \,.
    }

Up to the first-order in slow-varying parameters, the mode solutions
with proper initial conditions are (See Appendix \ref{appsec_mode}
for details)
    \ea{{\label{free_mode}}
        u_{\sigma}\left(\eta,k\right) &= \frac{\sqrt{\pi}}{2} \lrb{1+\frac{s_a}{2}} e^{\frac{i\pi}{2}\lrb{\nu_{\sigma}+\frac{1}{2}}} \sqrt{-\eta}  H^{(1)}_{\nu_{\sigma}}\lrb{(1+s_{a})x} \,,\\
        u_{s}\left(\eta,k\right) &= \frac{\sqrt{\pi}}{2}\left(1+\frac{s_{e}}{2}\right)e^{\frac{i\pi}{2}\left(\nu_{s}+\frac{1}{2}\right)}\sqrt{-\eta}H_{\nu_{s}}^{\left(1\right)}\left(\left(1+s_{e}\right)y\right) \,,
    }
with $x\equiv -c_a k\eta$ and $y\equiv -c_e k \eta$, and
\ea{{\label{nu}}
    \nu_{\sigma} &= \frac{3}{2} + \frac{1}{2} \lrb{ 2\epsilon +\eta_{\epsilon} +s_a }  \,,\\
    \nu_s & = \tilde{\nu}+\eta_{p}\frac{3}{4\tilde{\nu}}+s_{e}\tilde{\nu}+\epsilon\left(\tilde{\nu}-\frac{3}{4\tilde{\nu}}\right)
    \,,
} where $\tilde{\nu} =
\sqrt{\frac{9}{4}-\frac{\mu_{s}^{2}}{H^{2}}}$, $H^{(1)}$ is the
Hankel function of the first kind.

The ``decoupled" two-point functions for $\tilde{Q}_{\sigma}$ and
$\tilde{Q}_s$ are defined as
    \ea{{\label{def_tilde_GF}}
        \lrab{ \tilde{Q}_{\sigma}(\vk_1,\eta_{1}) \tilde{Q}_{\sigma} (\vk_2,\eta_{2}) }^{(0)} = (2\pi)^3 \delta^2 (\vk_1 + \vk_2)
        \tilde{G}_{k_1}(\eta_1,\eta_2) \,,\\
        \lrab{ \tilde{Q}_s (\vk_1,\eta_{1}) \tilde{Q}_s(\vk_2,\eta_{2}) }^{(0)} = (2\pi)^3 \delta^2 (\vk_1 + \vk_2)
        \tilde{F}_{k_1}(\eta_1,\eta_2) \,,\\
    }
where the supercript ``${}^{(0)}$" means in evaluating the above
expressions the coupling between adiabatic and entropy modes are
neglected, and
    \eq{{\label{decoupled_gf}}
        \tilde{G}_k(\eta_1,\eta_2) \equiv u_{\sigma}(\eta_1,k) u^{\ast}_{\sigma}(\eta_2,k)
        \,,\qquad \qquad \tilde{F}_{k}(\eta_1,\eta_2) \equiv u_s(\eta_1,k)
        u^{\ast}_s(\eta_2,k) \,,
    }
where $u_{\sigma}$, $u_s$ are given in (\ref{free_mode}), and
${}^\ast$ denotes complex conjugate.

In comoving gauge, the perturbation $Q_{\sigma}$ is directly related
to the three-dimensional curvature of  the constant time space-like
hypersurfaces. This gives the gauge-invariant quantity referred to
 the well-known ``comoving curvature perturbation":
    \eq{{\label{def_R}}
        \mathcal{R} \equiv \frac{H}{\dot{\sigma}} Q_{\sigma} \,,
    }
where $\dot{\sigma}$ is defined in (\ref{dot_sigma}). The entropy
perturbation $Q_s$ is automatically gauge-invariant by construction.
In practise, it is also convenient to introduce a renormalized
``isocurvature perturbation" defined by{\footnote{There is an
ambiguity in normalizing the entropy perturbation. Traditionally one
can choose the normalization condition to ensure that
$\mathcal{P}_{\mathcal{R}\ast} = \mathcal{P}_{\mathcal{S}\ast}$ when
modes cross the Hubble horizon. Our choice (\ref{def_S}) corresponds
to $\mathcal{P}_{\mathcal{R}\ast} / \mathcal{P}_{\mathcal{S}\ast}
\simeq c_e/c_a$, i.e the ratio of the speeds of sound of
isocurvature and curvature perturbations. Here
$\mathcal{P}_{\mathcal{R}\ast} \equiv
\mathcal{P}_{\mathcal{R}}(x_{\ast})$ and
$\mathcal{P}_{\mathcal{S}\ast} \equiv
\mathcal{P}_{\mathcal{S}}(y_\ast) $, see (\ref{exact_free_P}). }}
    \eq{{\label{def_S}}
        \mathcal{S} \equiv \frac{H}{\dot{\sigma}} Q_s \,.
    }
It is thus well-known result that the power spectra for curvature
perturbation and isocurvature perturbation around their respective
sound horizon-crossings  are (up to the first-order in slow-varying
parameters)
    \ea{{\label{exact_free_P}}
        \mathcal{P}_{\mathcal{R}}^{(0)}(x_{\ast}) &=
        \left.
        \bar{\mathcal{P}}_{\mathcal{R}}\left(1-2\epsilon-2s_{a}\right)F_{\nu_{\sigma}}\left(\left(1+s_{a}\right)x\right)
        \right|_{{c_ak}/{aH}=1} \,,
        \\
        \mathcal{P}_{\mathcal{S}}^{(0)}(y_{\ast}) &=
        \left.
        \bar{\mathcal{P}}_{\mathcal{S}}\left(1-2\epsilon-2s_{e}\right)F_{\nu_{s}}\left(\left(1+s_{e}\right)y\right)
        \right|_{{c_e k }/{a H}=1} \,,
    }
respectively, where the various parameters are defined in
(\ref{sv_para}) and $\bar{\mathcal{P}}_{\mathcal{R}} \equiv
\lrb{\frac{H}{2\pi}}^2\frac{1}{2\epsilon c_a}$ and
$\bar{\mathcal{P}}_{\mathcal{S}} \equiv
\lrb{\frac{H}{2\pi}}^2\frac{1}{2\epsilon c_e}$ are asymptotic
 values for the power spectra on superhorizon scales, and
    \eq{
        F_{\nu}(x) \equiv \frac{\pi}{2}x^{3}
        \left|H_{\nu}^{\left(1\right)}\left(x
        \right)\right|^{2} \,.
    }
 In (\ref{exact_free_P}), quantities on the right-hand-side of the equations
 are evaluated at the time of adiabatic or entropy sound horizon-crossings, i.e. ${c_ak}/{aH} = 1$ or ${c_e k}/{aH}=1$,
 respectively.
  In general since $c_a \neq c_e$, adiabatic and entropy modes cross their respective sound horizons at different times.
For later convenience, we introduce $x \equiv -c_a k \eta$ and
$y\equiv -c_e k \eta$, and
 in (\ref{exact_free_P}), $x_{\ast}$ and $y_{\ast}$ are their respective values around sound
 horizon-crossings,
  up to the first-order in slow-roll parameters which read
    \ea{
        x_{\ast} & \equiv \left. - c_a k \eta
        \right|_{c_ak/aH=1} \simeq 1 + \left. \epsilon\right|_{c_ak/aH=1} \,,\\
        y_{\ast} & \equiv \left. - c_e k \eta
        \right|_{c_ek/aH=1} \simeq  1 + \left. \epsilon\right|_{c_ek/aH=1}
        \,,
    }
where $\epsilon$ is the slow-roll parameter define in
(\ref{def_epsilon}).

 In the above derivation, $\mu_s^2/H^2$ and thus $3-2\nu_{s}$
are not supposed to be small. While in the following discussion, we
assume that $3-2\nu_{s}$ is of order $\sim\mathcal{O}(\epsilon)$.
Actually if the perturbation mode has an effective mass comparable
to the Hubble scale $H$, its quantum fluctuations on wavelengths
larger than the effective Compton wavelength would be suppressed,
then the system can be described effectively by a single-field.

\subsection{Cross-correlations}

Now we are at the point to evaluate the cross-power spectrum. As has
been stressed, the idea is treat the cross-coupling terms as
interaction vertices. From (\ref{Hc}), there are two types of
two-point cross-interaction vertices, as depicted in
fig.\ref{fig_cross_vertices}.
\begin{figure}[h]
        \centering
        \begin{minipage}{0.9\textwidth}
        \centering
            \begin{minipage}{0.44\textwidth}
                \centering
                \includegraphics[width=6cm]{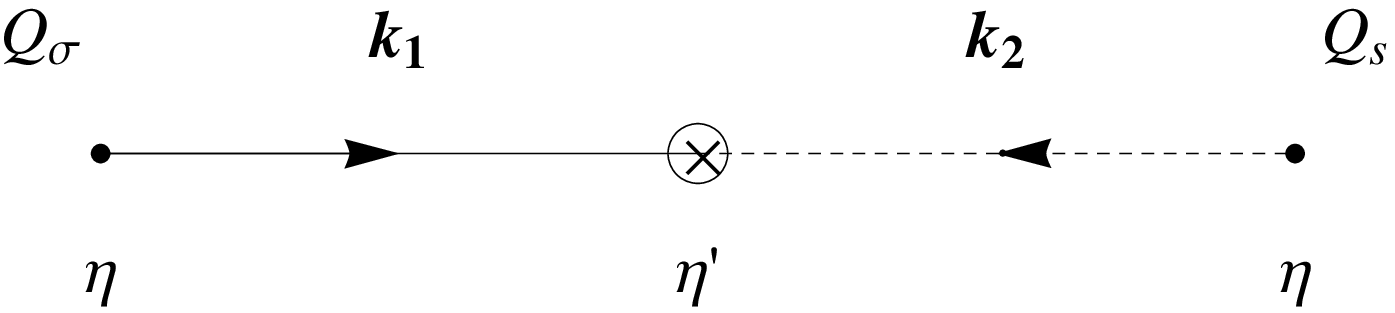}
            \end{minipage}
            \begin{minipage}{0.44\textwidth}
                \centering
                \includegraphics[width=6cm]{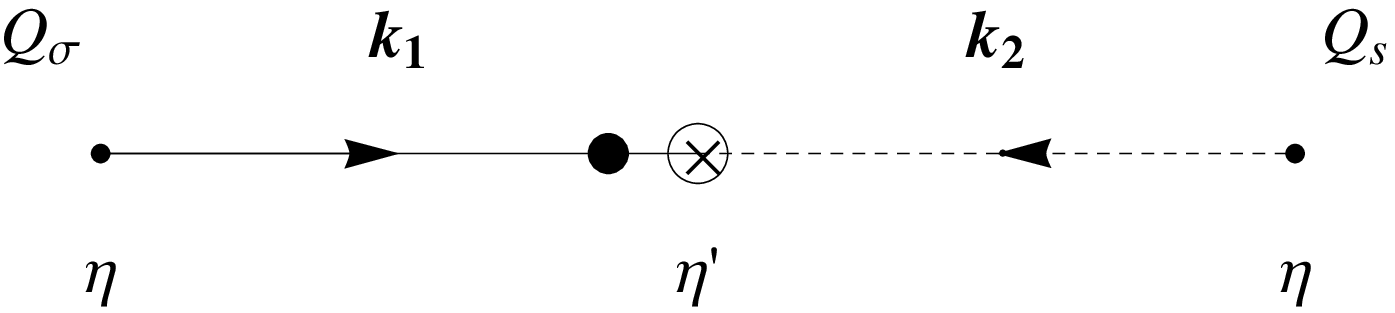}
            \end{minipage}
            \caption{Diagrammatic representations for cross-correlation vertices. Black line denotes adiabatic mode and dashed line denotes entropy mode. A ``$\otimes$" denotes cross-interaction vertex. There are two type of cross-interaction vertices.  A black dot denotes temporal derivative.}
            \label{fig_cross_vertices}
    \end{minipage}
    \end{figure}
In cosmological context, perturbative calculations of the
correlation functions are based on the ``in-in" formalism (see
Appendix \ref{appsec_inin} for a brief review). The leading-order
cross-correlation involves one cross-interaction vertex (see
fig.\ref{fig_cross_vertices}):
    \ea{{\label{cross_ex}}
        \left\langle \tilde{Q}_{\sigma}\left(\eta,\bm{k}_{1}\right)\tilde{Q}_{s}\left(\eta,\bm{k}_{2}\right)\right\rangle
        &= -2\Re \, i\int_{-\infty}^{\eta}d\eta'\left\langle \tilde{Q}_{\sigma}\left(\eta,\bm{k}_{1}\right)\tilde{Q}_{s}
        \left(\eta,\bm{k}_{2}\right)H_{c}\left(\eta'\right)\right\rangle
         \\
        &= -2\Re \, i\int_{-\infty}^{\eta} d\eta'\left[\mathcal{H}\xi\left(\eta'\right)\frac{d}{d\eta'}\tilde{G}_{k_{1}}
        \left(\eta,\eta'\right)\tilde{F}_{k_{1}}\left(\eta,\eta'\right)-\frac{\mathcal{H}z'\xi\left(\eta'\right)}{z}
        \tilde{G}_{k_{1}}\left(\eta,\eta'\right)\tilde{F}_{k_{1}}\left(\eta,\eta'\right)\right]\,.
    }
where $\tilde{G}$, $\tilde{F}$ are defined in (\ref{def_tilde_GF}).
From now on, we take the massless limit (i.e. $\nu_{\sigma} =\nu_s =
\frac{3}{2}$) for the decoupled two-point Green's functions
$\tilde{G}$ and $\tilde{F}$ defined in
(\ref{free_mode})-(\ref{decoupled_gf}) and also treat $H$ etc. as
constant in  evaluating the cross-correlation and the corrections to
adiabatic/entropy spectra. This is because not only that the exact
Green's functions are rather difficult to deal with analytically,
but also that the differences between using the massless Green's
function and using the exact Green's functions for our calculations
are higher-order in slow-roll parameters. More precisely, we thus
treat $\xi$ on the same footing as the other slow-varying
parameters, and identifies terms such as $\epsilon \xi$ as
higher-order quantities which can be neglected. At the end of our
calculations, the time-dependence of various parameters such as $H$
should be taken into account in order to get the correct tilts of
the spectra.

The cross-power spectrum for $Q_{\sigma}$ and $Q_s$ can be defined
as
    \eq{
        \left\langle Q_{\sigma}\left(\eta,\bm{k}_{1}\right) Q_{s}\left(\eta,\bm{k}_{2}\right)\right\rangle
        = \frac{c_{a} c_{e} }{a^2(\eta)} \left\langle \tilde{Q}_{\sigma}\left(\eta,\bm{k}_{1}\right)\tilde{Q}_{s}
        \left(\eta,\bm{k}_{2}\right)\right\rangle \equiv \left(2\pi\right)^{3}\delta^{3}\left(\bm{k}_{1}+\bm{k}_{2}\right) C_{\sigma
        s}\left(\eta,k_{1}\right) \,,
    }
From (\ref{cross_ex}) and (\ref{free_mode})-(\ref{decoupled_gf}),
after a straightforward calculation, $C_{\sigma s}$ can be written
in the form
 \eq{{\label{def_C}}
        C_{\sigma s}(\eta,k) =
        \frac{H^{2}}{2c_{a}k^{3}} \, \xi\,  \Gamma_c(x,\lambda) \,,
 }
with $x \equiv -c_{a} k \eta$ again and\eq{
         \lambda \equiv \frac{c_e}{c_a} \,,
        }
is the ratio of sound speeds of entropy and adiabatic modes, and
    \ea{{\label{Gamma_c}}
        \Gamma_c(x,\lambda) & \equiv \frac{ \left(1+\lambda\right)}{2\lambda^{2}}
         \Big\{2-x\left[2\sin\left(\left(1+\lambda\right)x\right)\textrm{Ci}\left(\left(1+\lambda\right)x\right)
         +\cos\left(\left(1+\lambda\right)x\right)\left(\pi-2\textrm{Si}\left(\left(1+\lambda\right)x\right)\right)\right] \\
        &\qquad -\frac{1-\lambda x^{2}}{1+\lambda}\left[2\cos\left(\left(1+\lambda\right)x\right)\textrm{Ci}
        \left(\left(1+\lambda\right)x\right)-\sin\left(\left(1+\lambda\right)x\right)\left(\pi-2\textrm{Si}
        \left(\left(1+\lambda\right)x\right)\right)\right] \Big\}
        \,,
        }
 Here we keep the $\eta$-dependence in the expression
for $\Gamma_c$ explicitly. It is not only because that there are
inflation models where the perturbation spectra evolves quickly even
after horizon crossing and never reaches the asymptotic values on
superhorizon scales ($x\rightarrow 0$), but also allows us a more
precise estimate of the spectra around the horizon crossing. It is
also interesting to note that the cross-power spectrum depends
explicitly on the ratio of the sound speeds for adiabatic and
entropy modes $\lambda = c_e/c_a$. Especially, the factor $\Gamma_c$
depends only on the ratio $\lambda$, while not on $c_a$ or $c_e$
themselves. The dimensionless cross-power spectrum between
$\mathcal{R}$ and $\mathcal{S}$ is given by{\footnote{Our result can
be compared
 with (e.g.) Eq.(79) in \cite{Lalak:2007vi}. Actually one may verify
that when setting $c_a = c_e$, (\ref{C_RS}) reduces to Eq.(79) in
\cite{Lalak:2007vi}. See Appendix \ref{appsec_compare} for
details.}}
    \ea{{\label{C_RS}}
        \mathcal{C}_{\mathcal{R}\mathcal{S}}(\eta,k) &\equiv
        \frac{k^3}{2\pi^2} \lrb{\frac{H}{\dot{\sigma}}}^2 C_{\sigma
        s}(\eta,k) = \bar{\mathcal{P}}_{\mathcal{R}} \,\xi\, \Gamma_c(x,\lambda) \,,
    }
where $\bar{\mathcal{P}}_{\mathcal{R}} \equiv \frac{1}{2\epsilon
c_a}\lrb{\frac{H}{2\pi}}^2 $ is asymptotic value for the
dimensionless power spectrum for the comoving curvature perturbation
on superhorizon scales as before.

(\ref{def_C})-(\ref{Gamma_c}) and (\ref{C_RS}) are one of the main
results in this note. The key point is that, at leading-order the
cross-power spectrum is of order $\sim \xi$, however its amplitude
is determined by the factor $\Gamma_c(x,\lambda)$. The dependence of
$\Gamma_c$ on $x$ and $\lambda$ is depicted in
fig.\ref{fig_Gamma_c}. \begin{figure}[h] \centering
    \begin{minipage}{1\textwidth}
        \centering
            \begin{minipage}{0.45\textwidth}
                \centering
                \includegraphics[width=8cm]{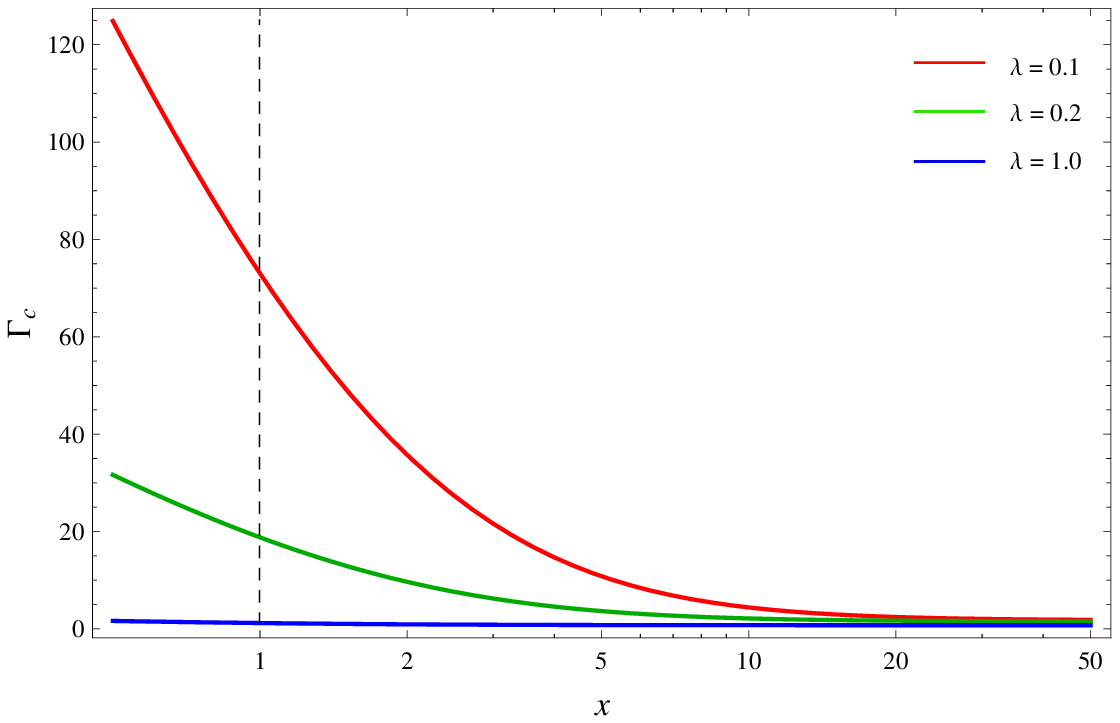}
            \end{minipage}
            \begin{minipage}{0.45\textwidth}
                \centering
                \includegraphics[width=8cm]{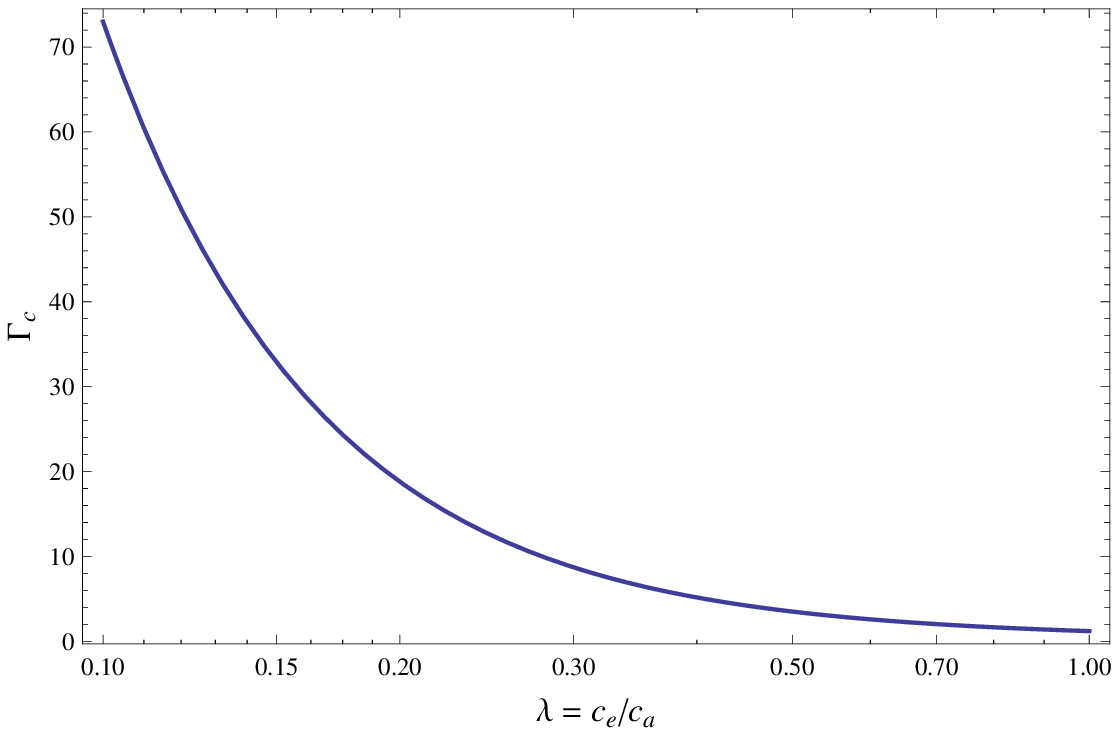}
            \end{minipage}
        \caption{$\Gamma_{c}$ as function of $x$ (left) and as function of $\lambda$ evaluated around adiabatic
        sound horizon-crossing at $x \approx 1$ (right).}
        \label{fig_Gamma_c}
        \end{minipage}
    \end{figure}

    Two comments are in order:
    \itm{
 \item From the left panel in fig.\ref{fig_Gamma_c}, when modes are deep
inside the horizons, $\Gamma_c \ll 1$ and the cross-spectrum are
indeed small. This confirms previous result that cross-correlations
among different perturbation modes are always negligible when modes
are deep in side their respective sound-horizons
\cite{Bartolo:2001vw,Byrnes:2006fr,Lalak:2007vi}, since there the
system reduces to a collection of weakly-coupled oscillators.
However, as firstly pointed out in \cite{Bartolo:2001vw}, as long as
the modes get closed to the horizon(s), the couplings and thus
cross-correlations among different modes become more and more
important, and the cross-correlations are generated when modes cross
their horizons. Our analysis also confirms this result. As depicted
in the left panel in fig.\ref{fig_Gamma_c}, when modes get closed to
the horizon, $\Gamma_c$ starts to increase, and its amplitude is
determined by $\lambda = c_e/c_a$, i.e. the ratio of the sound
speeds of isocurvature and curvature perturbations.

\item It is more interesting to note from the right panel in
fig.\ref{fig_Gamma_c} that, the value of $\Gamma_c$ and thus the
cross-correlation (around adiabatic sound horizon-crossing) can be
enhanced by small $\lambda$, i.e for models with $c_e \ll c_a$. This
phenomenon can be understood intuitively. As we know, the smaller
the sound speed is, the earlier the corresponding perturbation mode
crosses its sound horizon. The small $c_e/c_a$ ratio implies that
the isocurvature perturbation exits its sound horizon much earlier
before the curvature perturbation exits its sound horizon. Thus in
the process when curvature perturbation gets closed to the adiabatic
horizon, the isocurvature perturbation is already well outside its
entropic horizon and behaves as a nearly constant (rather than
highly oscillating) background, which acts as a nearly constant
source on the curvature perturbation. More precisely, the smaller
$\lambda$ is, the longer that the isocurvature perturbation behaves
as a source on the curvature perturbation, and the more significant
this ``accumulative" effect is. This fact causes the amplifications
of both cross-spectrum between curvature and isocurvature modes and
also the corrections to the spectra of curvature and isocurvature
perturbations by small $\lambda$, as we will show in the following
subsection. }

\subsection{Corrections to Spectra of Curvature and Isocurvature Perturbations}

Now we would like to investigate the leading-order corrections to
the power spectra of curvature and isocurvature perturbations, due
to the presence of the cross-interactions. It is interesting to note
that the leading-order corrections to $P_{\sigma}$ and $P_s$ from
the cross-interaction vertices involve two cross-interaction
vertices and thus are of $\sim \xi^2$, as depicted in
fig.\ref{fig_corr}.
\begin{figure}[h]
\centering
    \begin{minipage}{0.8\textwidth}
        \centering
            \begin{minipage}{0.46\textwidth}
                \centering
                \includegraphics[width=6cm]{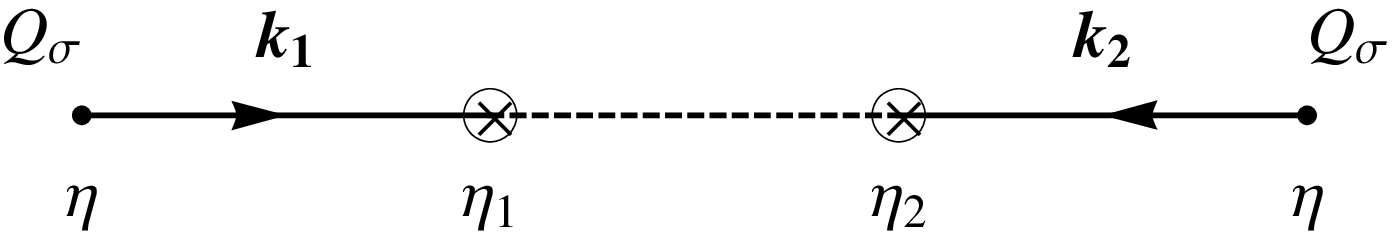}
            \end{minipage}
            \begin{minipage}{0.46\textwidth}
                \centering
                \includegraphics[width=6cm]{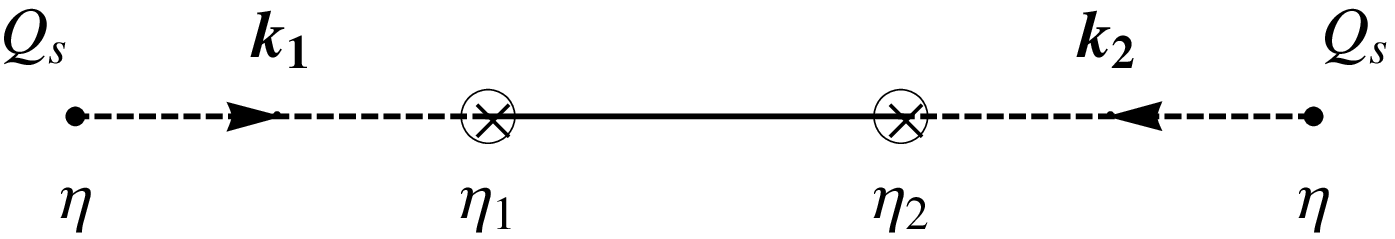}
            \end{minipage}
            \caption{Diagrammatic representations of the leading-order corrections to $P_{\sigma}$ and $P_s$. Recall that there are two types
            of cross-interaction vertices, and thus there are actually four different contributions, which we do not show here explicitly.}
        \label{fig_corr}
        \end{minipage}
    \end{figure}

The leading-order correction to the adiabatic power spectrum can be
denoted as
    \eq{
        \left\langle Q_{\sigma}\left(\eta,\bm{k}_{1}\right) Q_{\sigma}\left(\eta,\bm{k}_{2}\right)\right\rangle
        ^{(2)}\equiv\left(2\pi\right)^{3}\delta^{3}\left(\bm{k}_{1}+\bm{k}_{2}\right)P_{\sigma}^{(2)}\left(\eta,k_{1}\right)
        \,,
    }
here a superscript ``${}^{(2)}$" denotes that the contribution
involves two cross-interaction vertices, and
    \ea{{\label{P_sigma_2}}
        P_{\sigma}^{(2)}\left(\eta,k_{1}\right) = & -4\xi^{2}\times\Re\int_{-\infty}^{\eta}d\eta_{1}\int_{-\infty}^{\eta_{1}}d\eta_{2}F_{k_{1}}\left(\eta_{1},\eta_{2}\right)\\
        &\qquad \Big[ \frac{1}{\eta_{1}\eta_{2}}\frac{d}{d\eta_{1}}G_{k_{1}}\left(\eta,\eta_{1}\right)\frac{d}{d\eta_{2}}G_{k_{1}}\left(\eta,\eta_{2}\right)+\frac{1}{\eta_{1}\eta_{2}^{2}}\frac{d}{d\eta_{1}}G_{k_{1}}\left(\eta,\eta_{1}\right)G_{k_{1}}\left(\eta,\eta_{2}\right)\\
        &\qquad +\frac{1}{\eta_{1}^{2}\eta_{2}}G_{k_{1}}\left(\eta,\eta_{1}\right)\frac{d}{d\eta_{2}}G_{k_{1}}\left(\eta,\eta_{2}\right)+\frac{1}{\eta_{1}^{2}\eta_{2}^{2}}G_{k_{1}}\left(\eta,\eta_{1}\right)G_{k_{1}}\left(\eta,\eta_{2}\right) \Big]\\
        & +2\xi^{2}\int_{-\infty}^{\eta}d\eta_{1}\int_{-\infty}^{\eta}d\eta_{2}F_{k_{1}}\left(\eta_{1},\eta_{2}\right)\\
        &\qquad \times \Big[ \frac{1}{\eta_{1}\eta_{2}}\frac{d}{d\eta_{1}}G_{k_{1}}\left(\eta_{1},\eta\right)\frac{d}{d\eta_{2}}G_{k_{1}}\left(\eta,\eta_{2}\right)+\frac{1}{\eta_{1}\eta_{2}^{2}}\frac{d}{d\eta_{1}}G_{k_{1}}\left(\eta_{1},\eta\right)G_{k_{1}}\left(\eta,\eta_{2}\right)\\
        & \qquad
        +\frac{1}{\eta_{1}^{2}\eta_{2}}G_{k_{1}}\left(\eta_{1},\eta\right)\frac{d}{d\eta_{2}}G_{k_{1}}\left(\eta,\eta_{2}\right)+\frac{1}{\eta_{1}^{2}\eta_{2}^{2}}G_{k_{1}}\left(\eta_{1},\eta\right)G_{k_{1}}\left(\eta,\eta_{2}\right)
        \Big] \,.
        }
Here $G_k(\eta_1,\eta_2) \equiv
\frac{c_a^2}{a(\eta_1)a(\eta_2)}\tilde{G}_k(\eta_1,\eta_2)$ and
$F_k(\eta_1,\eta_2) \equiv
\frac{c_e^2}{a(\eta_1)a(\eta_2)}\tilde{F}_k(\eta_1,\eta_2)$. After a
straightforward calculation, $P_{\sigma}^{(2)}$ can be written in
the following form
    \eq{{\label{P_sigma_2}}
        P_{\sigma}^{(2)} =  \bar{P}_{\sigma} \xi^{2}\, \Gamma_{\sigma}\left(x,\lambda\right) \,,
    }
where $\bar{P}_{\sigma} \equiv \frac{H^2}{2c_a k^3} $, and
    \ea{
        \Gamma_{\sigma}\left(x,\lambda\right)
        = & \frac{1}{8\lambda^{3}x} \left\{x\left[-16(\lambda+1)\text{Ci}\left(2x\right)\cos\left(2x\right)
        +8(\lambda+1)\left(\pi-2\text{Si}\left(2x\right)\right)\sin\left(2x\right)+16\lambda+\pi^{2}+24\right]\right.\\
        &+x^{2}\left[16(\lambda+1)\text{Ci}\left(2x\right)\left(x\cos\left(2x\right)-2\sin\left(2x\right)\right)
        +x\left(\pi^{2}-8(\lambda+1)\left(\pi-2\text{Si}\left(2x\right)\right)\sin\left(2x\right)\right)\right.\\
        &\qquad \quad\left.-16(\lambda+1)\left(\pi-2\text{Si}\left(2x\right)\right)\cos\left(2x\right)
        -4\pi\cos\left(x(\lambda+1)\right)\right]\\
        &+4\left(x^{3}+x\right)\text{Ci}^{2}\left((\lambda+1)x\right)
        -8\left(x^{2}+1\right)\text{Ci}\left((\lambda+1)x\right)\sin\left(x(\lambda+1)\right)
        +4\left(x^{3}+x\right)\text{Si}^{2}\left((\lambda+1)x\right)\\&\left.+4\left(x^{2}+1\right)\text{Si}\left((\lambda+1)x\right)
        \left[2\cos\left(x(\lambda+1)\right)-\pi
        x\right]-4\pi\cos\left(x(\lambda+1)\right)\right\} + \frac{1}{\lambda^{3}}g_{\sigma}\left(x,\lambda\right)
        \,,
        }
with
    \eq{{\label{g_def}}
        g_{\sigma}\left(x,\lambda\right) \equiv
        \Re\left\{ \left(x+i\right)^{2}e^{i2x}
        \int_{x}^{+\infty} dz \frac{\left(z\lambda+i\right)e^{i\left(\lambda-1\right)z}}{z^{2}}
        \left[\text{Ei}\left(-iz\left(1+\lambda\right)\right)+i\pi\right]\right\}  \,.
    }

Similarly, the leading-order correction to entropy spectrum is
defined as
    \eq{
        \left\langle Q_{s}\left(\eta,\bm{k}_{1}\right) Q_{s}\left(\eta,\bm{k}_{2}\right)\right\rangle
        ^{(2)}\equiv\left(2\pi\right)^{3}\delta^{3}\left(\bm{k}_{1}+\bm{k}_{2}\right)P_{s}^{(2)}\left(\eta,k_{1}\right)
        \,,
    }
and we have
    \eq{
        P_{s}^{(2)} =  \bar{P}_{s} \xi^{2}\, \Gamma_{s}\left(y,\lambda\right) \,,
    }
where $\bar{P}_s\equiv \frac{H^2}{2c_e k^3}$ and $y\equiv -c_e k
\eta$, and
    \ea{
        \Gamma_{s}\left(y,\lambda\right)
        =& \frac{1}{8\lambda^{3}y} \left\{ y\left[4\left(y^{2}+1\right)\text{Ci}^{2}\left(\left(1+\lambda^{-1}\right)y\right)
        +4\left(y^{2}+1\right)\text{Si}^{2}\left(\left(1+\lambda^{-1}\right)y\right)-4\pi\left(y^{2}+1\right)
        \text{Si}\left(\left(1+\lambda^{-1}\right)y\right)\right.\right.\\
        &\qquad\qquad \left.+\pi^{2}y^{2}+8\lambda^{2}+\pi^{2}\right]-8\left(y^{2}+1\right)\lambda\text{Ci}\left(\left(1+\lambda^{-1}\right)
        y\right)\sin\left(y\left(\lambda^{-1}+1\right)\right)\\
        &\qquad\quad
        \left.-4\left(y^{2}+1\right)\lambda\left[\pi-2\text{Si}\left(\left(1+\lambda^{-1}\right)y\right)\right]
        \cos\left(y\left(\lambda^{-1}+1\right)\right)\right\} + \frac{1}{\lambda^{2}}g_{s}\left(y,\lambda\right)
        \,,
    }
with
    \eq{
        g_s \left(y,\lambda\right) \equiv \Re\left\{
        \left(y+i\right)^{2}e^{i2y}
        \int_{y}^{+\infty} dz e^{i\left(\lambda^{-1}-1\right)z}\frac{i-z}{z^{2}}
        \left[\text{Ei}\left(-iz\left(1+\lambda^{-1}\right)\right)+i\pi\right]\right\}
        \,.
        }

The $x$, $y$ and $\lambda$-dependence of $\Gamma_{\sigma}$ and
$\Gamma_s$ are depicted in fig.\ref{fig_Gamma_sigma} and
fig.\ref{fig_Gamma_s}.
\begin{figure}[h] \centering
    \begin{minipage}{1\textwidth}
        \centering
            \begin{minipage}{0.45\textwidth}
                \centering
                \includegraphics[width=7cm]{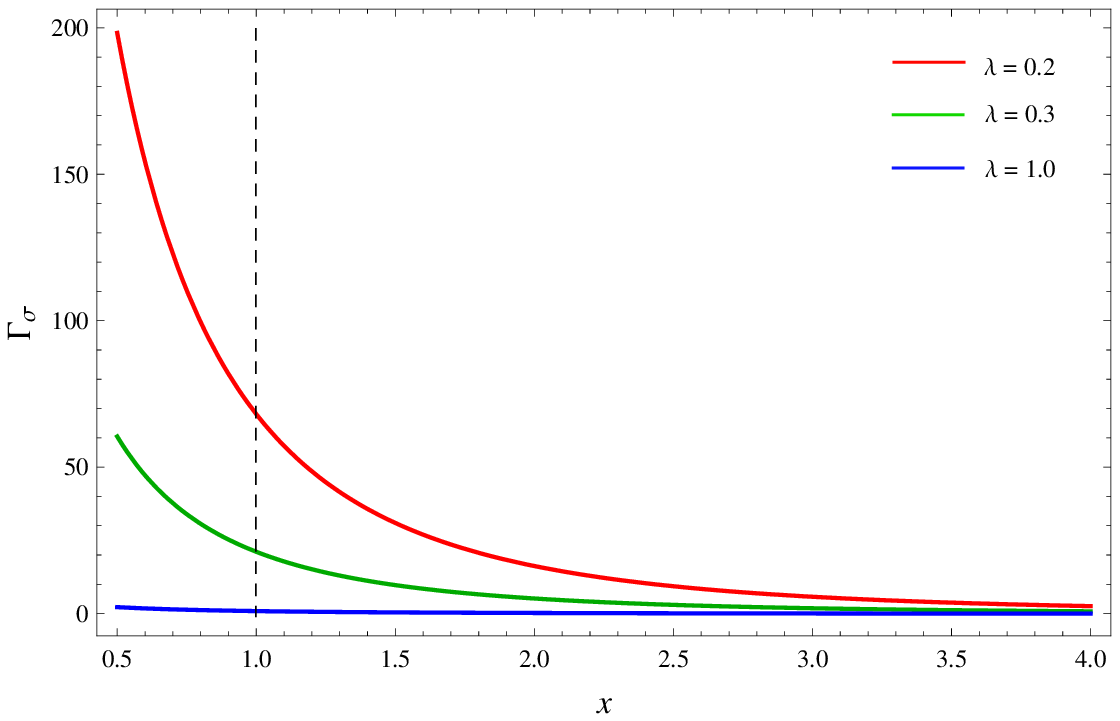}
            \end{minipage}
            \begin{minipage}{0.45\textwidth}
                \centering
                \includegraphics[width=7cm]{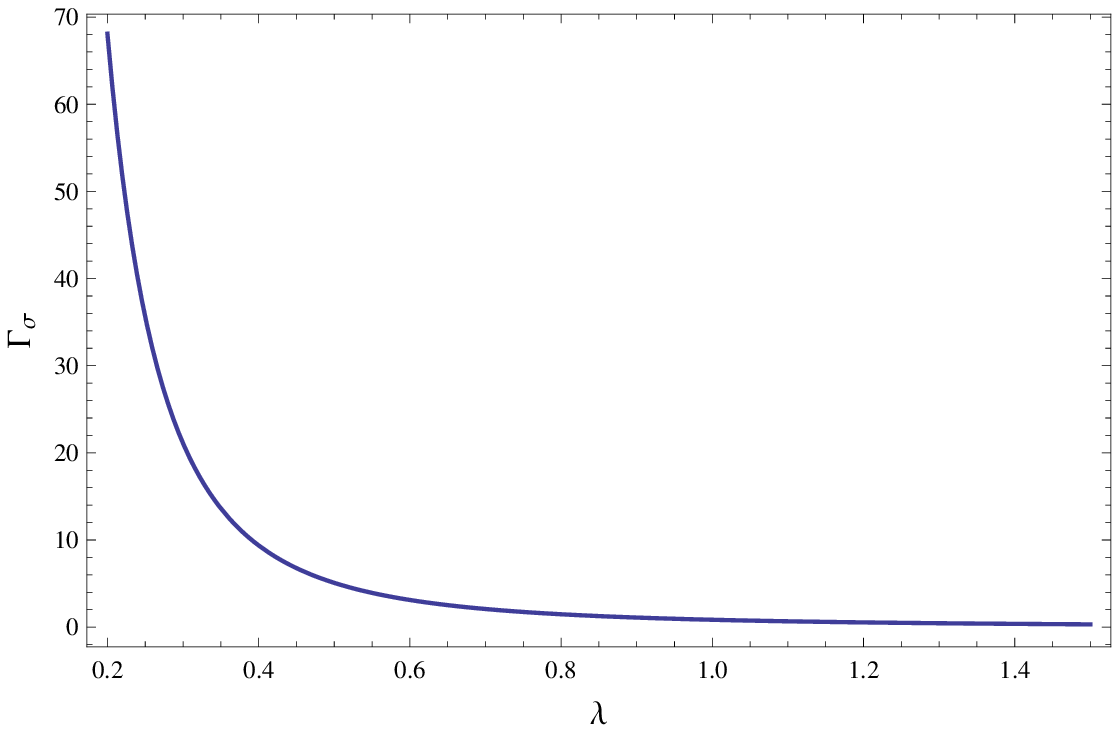}
            \end{minipage}
        \caption{$\Gamma_{\sigma}$ as function of $x$ (left) and as function of $\lambda$ evaluated
        around the adiabatic
        sound horizon-crossing at $x \approx 1$ (right).}
        \label{fig_Gamma_sigma}
        \end{minipage}
    \end{figure}

\begin{figure}[h] \centering
    \begin{minipage}{1\textwidth}
        \centering
            \begin{minipage}{0.45\textwidth}
                \centering
                \includegraphics[width=7cm]{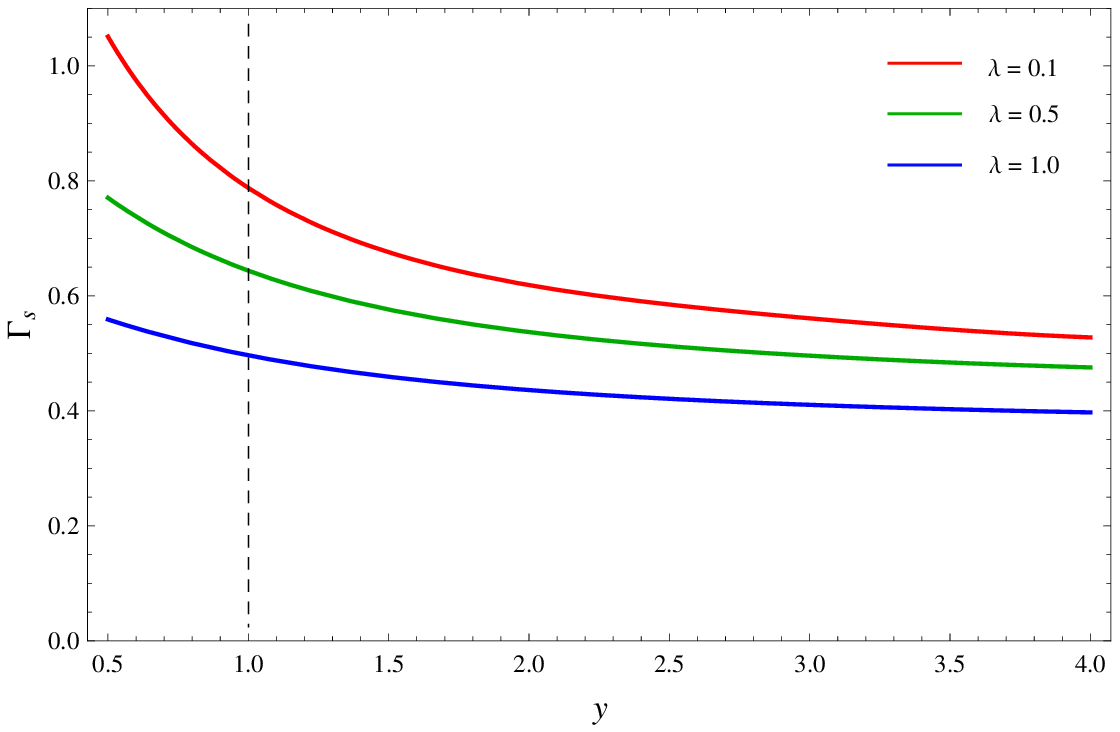}
            \end{minipage}
            \begin{minipage}{0.45\textwidth}
                \centering
                \includegraphics[width=7cm]{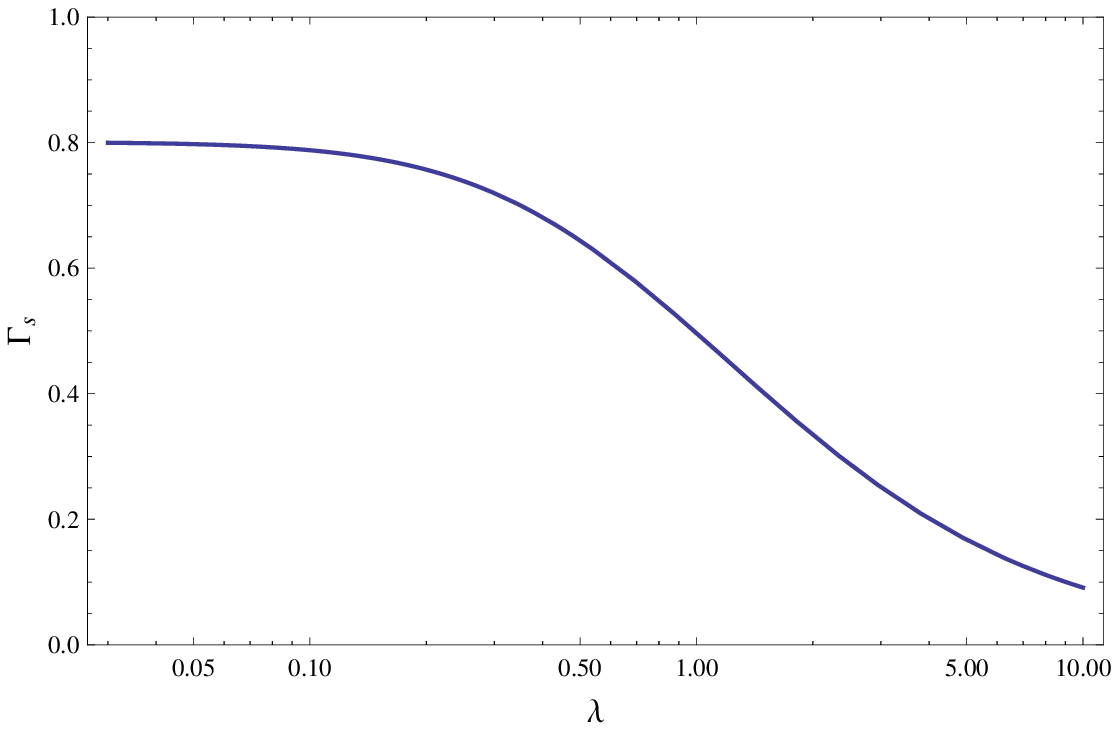}
            \end{minipage}
        \caption{$\Gamma_{s}$ as function of $y$ (left) and as function of $\lambda$ evaluated around the entropy
        sound horizon-crossing at $y \approx 1$ (right).}
        \label{fig_Gamma_s}
        \end{minipage}
    \end{figure}

Several comments are in order:
    \itm{
        \item The left figures in fig.\ref{fig_Gamma_sigma} and fig.\ref{fig_Gamma_s} show $\Gamma_{\sigma}$ or $\Gamma_s$ as
        functions
        of $x$ or $y$ respectively, for different values of $\lambda$. We take $\Gamma_{\sigma}$ as example.
        Since $x \equiv -c_a k \eta$, it implies that
        for mode with fixed $k$, when deep in the sound horizon ($x\gg 1$),
        corrections to the power spectrum from the cross-interactions are
        small. Again, this verifies previous argument that adiabatic and entropy perturbations can be treated as decoupled when they are deep
        inside
        the horizon, while when modes approach the horizon, i.e. $x\approx
        1$, the modification to the power spectrum starts to
        increase \cite{Bartolo:2001vw}.
         The conclusion is the same for $\Gamma_s$.


        \item The most interesting point is that, as for the cross-correlation between adiabatic and entropy modes,
        the strength of the corrections to adiabatic/entropy spectra due to the presence of cross-couplings are also determined by the
        parameter $\lambda \equiv c_e/c_a$, i.e. the ratio of the sound speeds for entropy and adiabatic modes.
        It has been known that for $c_a=c_e$ i.e. $\lambda=1$, the cross-correlation and
        also the corrections to the ``decoupled" spectra are proportional to the cross-coupling and thus are expected to be small \cite{Bartolo:2001vw,Byrnes:2006fr,Lalak:2007vi}.
         However, from the
        right figures in fig.\ref{fig_Gamma_sigma} and fig.\ref{fig_Gamma_s},
        this
        corrections
        can be enhanced by small $\lambda$, that is for models with $c_e \ll
        c_a$. Especially, this enhancement is most significant for the adiabatic mode
(fig.\ref{fig_Gamma_sigma}). As explained before,
        when $c_e \ll c_a$ the entropy mode cross the horizon much earlier than the adiabatic mode,
        and thus act as a nearly constant (rather than highly oscillating) source  on the evolution of adiabatic
        mode. Thus the smaller $\lambda$ is, the longer that the entropy mode behaves as a source, and
         the more significant this accumulative effect is.

        \item Inversely, this enhancement is not significant for
        the entropy mode. Actually from the right panel in fig.\ref{fig_Gamma_s}, when $\lambda
        >
        1$ i.e. $c_e > c_a$, the correction to the power spectrum
        of entropy mode is  suppressed when $\lambda$ goes
        large. Intuitively, this is because that, as is well-known, on super-horizon
        scales entropy modes can act as sources for the evolution of adiabatic mode, while
        inversely adiabatic mode can never act as a source for entropy
        modes. Thus, the larger $\lambda$ is, the earlier the adiabatic mode exits its
        horizon, and the less it affects the evolution of entropy
        modes. As an extremal case, one can verify that $\Gamma_s \rightarrow 0$ when $\lambda \rightarrow
        \infty$, which implies that in this case the entropy mode is not affected by adiabatic mode and evolves
        freely. Moreover from the right panel in fig.\ref{fig_Gamma_s}, when $\lambda \rightarrow 0$, $\Gamma_s$
        approaches a constant value rather than  blowing up, since
        in this case the adiabatic mode which is highly
        oscillating affects the entropy mode with a nearly constant
        strength.
    }

What we are eventually interested in are the spectra of the
curvature and isocurvature perturbations, which are defined in
(\ref{def_R}) and (\ref{def_S}).  After using (\ref{exact_free_P}),
the power spectrum for the curvature perturbation, including the
leading-order corrections from the cross-interactions, and also
including the corrections to both Green's function and Hubble
parameter in first-order slow-roll parameters, takes the form
    \ea{{\label{PR_final}}
        \mathcal{P}_{\mathcal{R}}(x) &=
        \mathcal{P}^{(0)}_{\mathcal{R}}(x) +
        \mathcal{P}^{(2)}_{\mathcal{R}}(x) \\
        &\simeq
        \bar{\mathcal{P}}_{\mathcal{R}}\left(1+x^{2}\right)\left[1-2\epsilon-2s_{a}
        +\left(\nu_{\sigma}-\frac{3}{2}\right)f\left(\left(1+s_{a}\right)x\right)+s_{a}\frac{2x^{2}}{1+x^{2}}+\frac{\xi^{2}}{\left(1+x^{2}\right)}\Gamma_{\sigma}\left(x,\lambda\right)\right]
        \,,
        }
where
    \eq{{\label{def_f}}
        f(x) \equiv \frac{\pi
        x^{3}}{2\left(1+x^{2}\right)}\left(\frac{\partial}{\partial\nu}\left|H_{\nu}^{\left(1\right)}\left(x\right)\right|^{2}\right)_{\nu=\frac{3}{2}}
        \,.
    }
In deriving the above expression, (\ref{app_def_f}) and
(\ref{app_def_f}) are used.
 Similarly, the power spectrum for isocurvature perturbation is
    \eq{{\label{PS_final}}
        \mathcal{P}_{\mathcal{S}}(y) \simeq
        \bar{\mathcal{P}}_{\mathcal{S}}\left(1+y^{2}\right)\left[1-2\epsilon-2s_{e}
        +\left(\nu_{s}-\frac{3}{2}\right)f\left(\left(1+s_{e}\right)y\right)+s_{e}\frac{2y^{2}}{1+y^{2}}
        +\frac{\xi^{2}}{\left(1+y^{2}\right)}\Gamma_{s}\left(y,\lambda\right)\right]
        \,.
    }
(\ref{C_RS}), (\ref{PR_final}) and (\ref{PS_final}) are the main
results in this note, and can be viewed as generalizations of
previous results, e.g. Eq.(78)-(80) in \cite{Lalak:2007vi}.

\subsubsection{Deep Inside the Horizon}

In \cite{Bartolo:2001vw}, an oscillating mechanism was introduced to
study the cross-correlations between perturbations, where it was
found that when deep inside the Hubble horizon different modes
evolve independently and can be considered as good mass eigenstates,
thus the cross-correlations are indeed small. Intuitively, when deep
inside the horizon, the system become weakly-coupled oscillators in
Minkowski background in which the couplings among them are assumed
to be small (of order slow-roll parameters). In fact, as an explicit
confirmation, one can show that
    \eq{{\label{Gamma_c_ih}}
        \Gamma_c(x,\lambda ) \quad \xrightarrow[]{~~ x \gg 1
        ~~} \quad \frac{2+\lambda }{(1+\lambda )^2} \,,
    }
which is independent of $x$. Similarly, it can be verified that
$g_{\sigma,s}(x,\lambda)$ approach constant values (independent of
$x$) when $x \gg 1$,
    \[
        g_{\sigma}(x,\lambda) \rightarrow
        \frac{\lambda}{2\left(1+\lambda\right)} \,,\qquad g_{s}(y,\lambda) \rightarrow -\frac{\lambda}{2(1+\lambda)} \,,
    \]
thus
    \ea{{\label{Gamma_sigma_s_ih}}
        \Gamma_{\sigma}(x,\lambda) \quad
        &\xrightarrow[]{~~x\gg 1 ~~} \quad  -\frac{1}{2 (1+\lambda )^2}
        \,,\\
        \Gamma_{s}(y,\lambda) \quad
        &\xrightarrow[]{~~y\gg 1 ~~} \quad  \frac{1+2\lambda}{2(1+\lambda)^{2}}  \,.
    }
From (\ref{Gamma_c_ih}) and (\ref{Gamma_sigma_s_ih}), it is obvious
that
    \eq{
        \max \Gamma_i \sim \mathcal{O}(1) \,, \qquad i=c,\sigma,s
        }
thus we can conclude that when deep inside the horizon, the
cross-correlation between curvature and isocurvature perturbation is
always smaller than $\mathcal{O}(\xi)$, and the corrections to
curvature/isocurvature perturbation due to this cross-interactions
are always smaller than $\mathcal{O}(\xi^2)$. This confirms previous
investigation that the couplings between adiabatic/entropy
perturbations can be neglected when modes are sub-Hubble.

However, as was firstly pointed out in \cite{Bartolo:2001vw} and was
also analyzed in \cite{Byrnes:2006fr,Lalak:2007vi}, the most
important lesson we get is that,  the cross-interactions between
adiabatic and entropy modes at linear level{\footnote{When
investigating non-Gaussian features, higher-order interactions among
different field modes are important, e.g. in multi-field models or
in inflaton-curvaton models. Loop corrections also involve
cross-interactions, see e.g. \cite{Gao:2009fx}}}, which are
negligible when modes are deep inside the horizons, are generated
and amplified when modes cross their sound horizons. Our analysis in
this work also confirms this fact.

\subsubsection{Around Horizon-crossings}


As has been stressed before, (\ref{C_RS}), (\ref{PR_final}) and
(\ref{PS_final}) are the general expressions for power spectra for
curvature and isocurvature perturbations and their cross power. Thus
in general (\ref{C_RS}), (\ref{PR_final}) and (\ref{PS_final}) are
needed to evaluate more precisely the amplitude of the powers and
the spectral tilts around horizon crossing ($x_{\ast} \simeq 1+
\epsilon_a$ and $ y_{\ast} \simeq 1+\epsilon_e$). In this
subsection, for our purpose to get a glance of the effects of the
cross-correlations, we use superhorizon asymptotic limits of
(\ref{C_RS}), (\ref{PR_final}) and (\ref{PS_final}) to evaluate
various quantities around horizon-crossing{\footnote{This is the
traditional treatment in the literature. However, the spectral
indices of the spectra in multi-field models in general should be
evaluated by using the general expressions (\ref{C_RS}),
(\ref{PR_final}) and (\ref{PS_final}), rather than naively by using
the curvature perturbation as in single-field models. See e.g. the
numerical discussions in \cite{Lalak:2007vi}.}}.

 The cross-power spectrum around adiabatic
horizon-crossing is approximately{\footnote{(\ref{C_RS_hc}) can be
compared with (e.g.) Eq.(37) in \cite{Byrnes:2006fr}. If we set
$\lambda\equiv 1$ and $\xi = -2\eta_{\sigma s}$, (\ref{C_RS_hc})
reduces to Eq.(37) in \cite{Byrnes:2006fr}.}}
    \eq{{\label{C_RS_hc}}
        \left. \mathcal{C}_{\mathcal{R}\mathcal{S}} \right|_{\frac{c_ak}{aH}=1} \simeq \left.
        \bar{\mathcal{P}}_{\mathcal{R}}\xi C(\lambda) \right|_{\frac{c_ak}{aH}=1}\,,
        \qquad\qquad \textrm{with}\; C(\lambda)\equiv  \frac{1+\lambda-\ln(1+\lambda)-\gamma }{\lambda^{2}}
        \,,
    }
where $\gamma \approx 0.5772$ is the Euler-Mascheroni constant. The
spectral index is
    \ea{{\label{n_C}}
        \left. n_{\mathcal{C}} \right|_{\frac{c_ak}{aH}=1} &\equiv \left. \od{\ln \mathcal{C}_{\mathcal{R}\mathcal{S}}}{\ln
        k} \right|_{\frac{c_ak}{aH}=1}
        \simeq
        \left. \lrsb{ -2\epsilon-\eta+\eta_{\xi}+\tilde{C}
        s_{a}-\left(\tilde{C}+1\right)s_{e} } \right|_{\frac{c_ak}{aH}=1}
        \,,
        }
with $\tilde{C}\equiv 1-1/(\left(\lambda+1\right) C (\lambda) )$.
Note that in (\ref{C_RS_hc}) and (\ref{n_C}) all quantities are
evaluated at the time of adiabatic sound horizon-crossing, i.e.
$c_ak/aH=1$.
  From (\ref{C_RS_hc}) it is
explicit that up to first-order in the cross-interaction coupling
$\xi$, the cross-power spectrum is $\sim C(\lambda)\xi$, it can be
enhanced by small $\lambda$ due to the factor $C(\lambda)$, which
scales as $1/\lambda^2$ for small $\lambda$. Note that for
$\lambda=1$ i.e. for $c_a = c_e$, $C(\lambda)$ reduces to the
familiar value $C(1)=2-\ln 2-\gamma \approx 0.7296$. Thus our
calculations can be viewed as generalization of previous results in
\cite{Byrnes:2006fr,Lalak:2007vi}.

Similarly, power spectra for the curvature and isocurvature
perturbations around their respective horizon-crossings are
    \ea{{\label{spectra_hc}}
        \left. \mathcal{P}_{\mathcal{R}} \right|_{\frac{c_ak}{aH}=1}
        &\simeq \left. \bar{\mathcal{P}}_{\mathcal{R}}\left(1-2\epsilon-2s_{a}\right)
        \left[1+\left(3-2\nu_{\sigma}\right)\left(\gamma-2+\ln2\right) + \xi^2 a_{\sigma}(\lambda) \right] \right|_{\frac{c_ak}{aH}=1} \,,\\
        \left. \mathcal{P}_{\mathcal{S}} \right|_{\frac{c_ek}{aH}=1}
        &\simeq \left. \bar{\mathcal{P}}_{\mathcal{S}}\left(1-2\epsilon-2s_{e}\right)
        \left[1+\left(3-2\nu_{e}\right)\left(\gamma-2+\ln2\right) + \xi^2 a_{s}(\lambda) \right] \right|_{\frac{c_ek}{aH}=1} \,,\\
    }
where $a_{\sigma,s}(\lambda)$ are functions of $\lambda$ only. In
getting (\ref{spectra_hc}), we have neglected all terms proportional
to (such as) $\epsilon \xi^2$ etc., i.e. we only keep the
leading-order contributions from $\xi^2$. For general $\lambda$, it
is difficult to abstract analytical expressions for
$a_{\sigma,s}(\lambda)$. Their numerical results are depicted in
fig.\ref{fig_a_lambda}, Form which it is explicit that the
corrections to both curvature and isocurvature power spectra can be
enhanced by small $\lambda$.
\begin{figure}[h]
\centering
    \begin{minipage}{1\textwidth}
        \centering
            \begin{minipage}{0.45\textwidth}
                \centering
                \includegraphics[width=7cm]{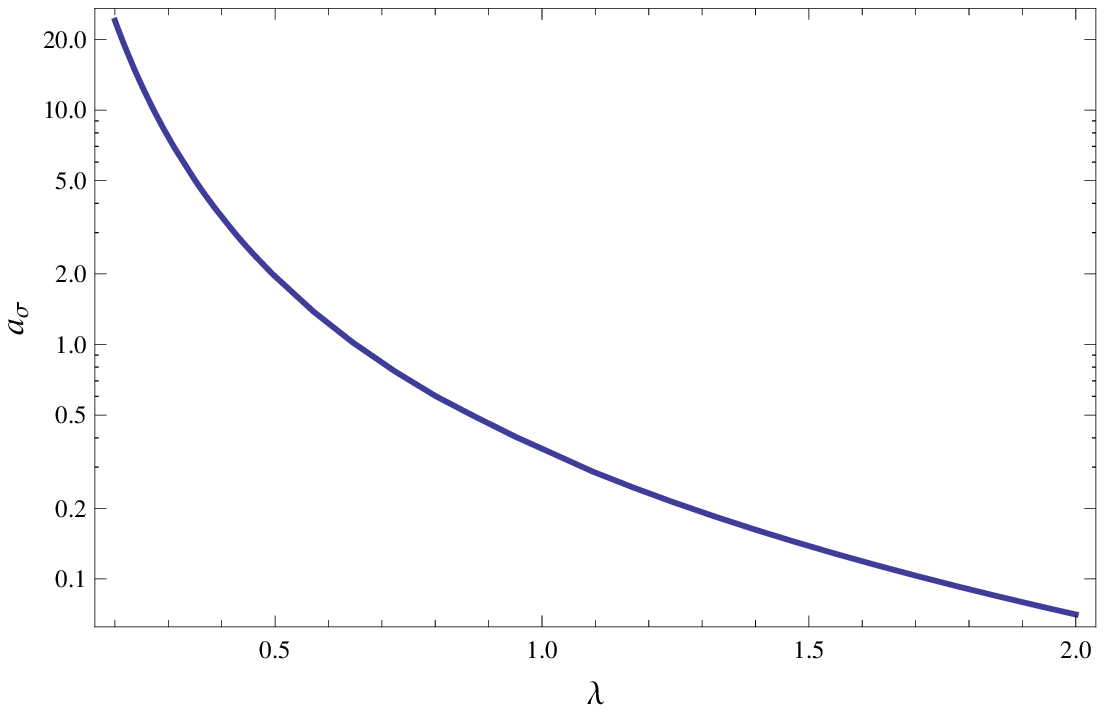}
            \end{minipage}
            \begin{minipage}{0.45\textwidth}
                \centering
                \includegraphics[width=7cm]{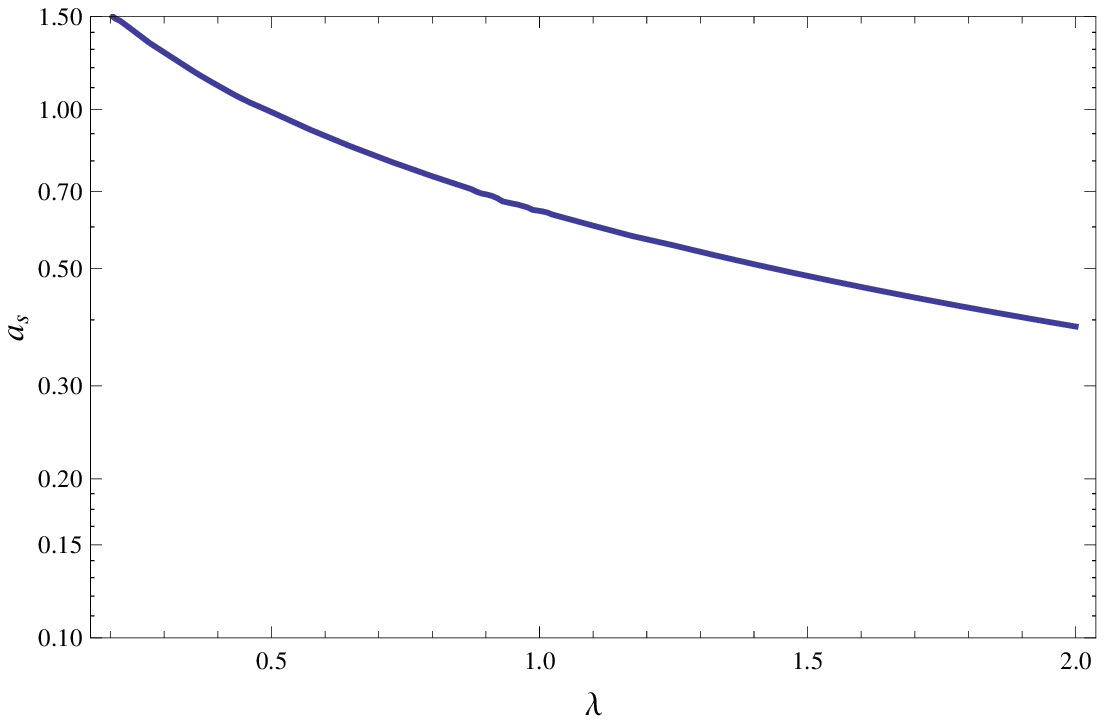}
            \end{minipage}
        \caption{Numerical evaluation for $a_{\sigma}(\lambda)$ and $a_s(\lambda)$ as functions of $\lambda$.}
        \label{fig_a_lambda}
        \end{minipage}
    \end{figure}

In practice, it is convenient to introduce a dimensionless
correlation angle $\Delta$ to characterize the strength of the
cross-correlation:
    \eq{
        \cos\Delta \equiv \frac{\mathcal{C}_{\mathcal{R}\mathcal{S}}}{\sqrt{ \mathcal{P}_{\mathcal{R}} \mathcal{P}_{\mathcal{S}}
        }} \,.
    }
From (\ref{C_RS}), (\ref{PR_final}) and (\ref{PS_final}), and using
the fact that around the adiabatic sound horizon-crossing
$\bar{\mathcal{P}}_{\mathcal{R}}/\bar{\mathcal{P}}_{\mathcal{S}}
\approx \lambda$, we get the correlation angle around adiabatic
sound horizon-crossing
    \eq{{\label{Delta_app}}
        \left. \cos\Delta \right|_{\frac{c_ak}{aH}=1} \approx
        \left. \xi\sqrt{\lambda}\Gamma_{c}\left(x,\lambda\right) \right|_{\frac{c_ak}{aH}=1}
        \,.
    }
In getting (\ref{Delta_app}) we take the approximation under the
assumption that the power spectra of curvature/isocurvature
perturbations are still dominated by their ``decoupled" values
$\mathcal{P}_{\mathcal{R}}^{(0)}$ and
$\mathcal{P}_{\mathcal{S}}^{(0)}$, i.e we assume that the
corrections to the spectra would not exceed their respective
decoupled values (they are indeed corrections). A more precise
evaluation can be done by using (\ref{C_RS_hc}) and
(\ref{spectra_hc}) and the corresponding correlation angle around
adiabatic horizon-crossing is depicted in fig.\ref{fig_angle}.
\begin{figure}[h] \centering
    \begin{minipage}{1\textwidth}
        \centering
            \begin{minipage}{0.45\textwidth}
                \centering
                \includegraphics[width=8cm]{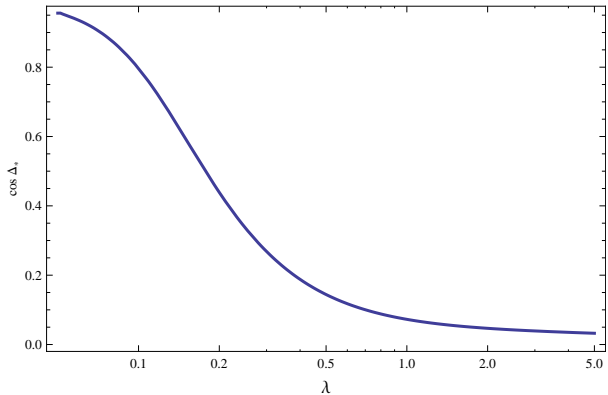}
            \end{minipage}
        \caption{The correlation angle  $\cos\Delta_{\ast}$ around adiabatic sound horizon-crossing as function of $\lambda$. Here $\xi$ is chosen as 0.1}
        \label{fig_angle}
        \end{minipage}
    \end{figure}
    It
immediately follows that, for $\lambda=1$ (the case for canonical
kinetic terms or multi-DBI case with $c_a=c_e=c_s$) or larger (a
concrete example is the case for multi-field $k$-inflation
\cite{Langlois:2008mn,Gao:2008dt}, where $c_a = c_s \ll 1$ and
$c_e=1$), the correlation angle is small and the cross-correlation
between curvature and isocurvature perturbations are suppressed.
However, as has been stressed before, for models with $c_e < c_a$ or
even $c_e \ll c_a$, the cross-correlations will be highly enhanced
by small $\lambda$. Thus, for models with $\xi$ much larger than the
slow-roll parameters which are of order $ 10^{-2}$ or with $c_e \ll
c_a$, the curvature and isocurvature perturbations are highly
correlated when exiting the adiabatic sound horizon. In this case,
the ``decoupled" power spectra $\mathcal{P}_{\mathcal{R}}^{(0)}$ and
$\mathcal{P}_{\mathcal{S}}^{(0)}$ are not good approximations, and
the cross-correlations between curvature and isocurvature
perturbations much be taken into account.


\section{Conclusion and Discussion}

In this paper we investigate the effects of cross-interactions
between curvature perturbation and isocurvature perturbations on
cross-power spectrum and the power spectra of curvature/isocurvature
perturbations themselves, before and around horizon crossing.
Previous investigations of the cross-correlations are based on
diagonalizting the coupled equations of motion
\cite{Bartolo:2001vw,Byrnes:2006fr,Lalak:2007vi}. However, one can
verify  that, for models with different adiabatic/entropy speeds of
sound, the ``diagonalization" cannot be done easily and thus the
treatments in \cite{Bartolo:2001vw,Byrnes:2006fr,Lalak:2007vi}
cannot be viewed as good approximation in this case. Thus, in this
work, the cross-couplings are taken as two-point interaction
vertices, and a field-theoretical perturbative approach is taken to
evaluate the cross-power spectrum etc{\footnote{This perturbative
method is standard, which has also been introduced independently by
Chen and Wang very recently in evaluating the ``transfer" between
adiabatic and entropy modes as well as their implications on
non-Gaussianities\cite{cw}.}}.

The main results in this work are summarized in (\ref{C_RS}),
(\ref{PR_final}) and (\ref{PS_final}). Our analysis confirms
previous conclusion that the cross-correlations, which can be safely
neglected when modes are deep inside the horizons, are generated
when modes cross their sound horizons \cite{Bartolo:2001vw}.
Moreover, the most interesting phenomenon get in this work is that
the cross-correlation (and also the corrections to power spectra of
curvature/isocurvature perturbations) can be enhanced by small
$c_e/c_a$ ratio, where $c_a$ and $c_e$ are the sound speeds of
curvature and isocurvature perturbations respectively. As has been
stressed before, this happens since in models with $c_e/c_a \ll 1$,
in the process curvature perturbation getting closed to the
adiabatic horizon, the isocurvature perturbation is already well
outside its entropic horizon and behaves as a nearly constant
(rather than highly oscillating) background, which acts as a nearly
constant source on the curvature perturbation. This fact causes the
amplifications of both cross-spectrum between curvature and
isocurvature modes and also the corrections to the spectra of
curvature and isocurvature perturbations.

To end this note, we would like to comment some limitations and also
possible extensions of the investigation in this note:
    \itm{
        \item The approach in evaluating the cross-correlation and
        the corrections to (decoupled) power spectra of curvature
        and isocurvature perturbation is perturbative. The coupling
        $\xi$ is assumed to be small. This is indeed the case for
        canonical kinetic terms, but in models (e.g.) considered in
        this note, $\xi$ may be not small enough for the
        perturbative approach to be valid.

        \item In general the cross-couplings between adiabatic and
entropy modes are complicated. In this note we make the assumption
that the cross-coupling $\xi$ in (\ref{Hamiltonian}) is slow-varying
in time. This is a simplification for abstracting the basic
properties of the cross-correlations, but definitely a more detailed
analysis of the cross-interactions is needed.

    \item Moreover, in this note we
only focus on the cross-correlations around the horizon-crossing. In
order to get the resulting primordial power spectra on large scales,
and to compare the predictions of a multi-inflation model with
observations, one must then solve the coupled system described by
the full equations of motion. In some particular case and within the
slow-roll approximation, one can arrive at an analytical expression
for the spectra on large scales, e.g. through the ``transfer matrix"
method \cite{Wands:2002bn}. In general, however, a numerical
approach is needed.

    \item As has been stressed before, (\ref{C_RS}), (\ref{PR_final}) and
(\ref{PS_final}) are the general expressions for power spectra for
curvature and isocurvature perturbations and their cross power. In
general, in multi-field models it may not be permitted to use the
later time (superhorizon limit) asymptotic expressions to evaluate
the powers around horizon-crossing, since there may be inflation
scenarios where the perturbation evolve quickly after horizon
crossing and never reach the asymptotic values. Thus in general
(\ref{C_RS}), (\ref{PR_final}) and (\ref{PS_final}) are needed to
evaluate more precisely the amplitude of the powers and also the
spectral indices around horizon-crossings ($\frac{c_ak}{aH}=1$ and
$\frac{c_e k}{aH}=1$ respectively).

    \item One possible application of the formalism and result in this work is that, as firstly pointed out
in \cite{Bartolo:2001cw}, one should expect the ``transfer" of
non-Gaussianities from entropy perturbations to the adiabatic
perturbation (see fig.\ref{fig_ng}), if the cross-correlations
between adiabatic mode and entropy modes are larger than $\sim
\mathcal{O}(\epsilon)$ and thus cannot be neglected. Especially,
there are inflation scenarios in which the non-linearities in
adiabatic mode itself are small, however the possible large
non-Gaussianities in entropy modes could transfer to the
non-Gaussianities in adiabatic mode through the cross-interactions.
\begin{figure}[h]
    \centering
    \begin{minipage}{0.6\textwidth}
    \centering
    \begin{minipage}{0.4\textwidth}
    \centering
    \includegraphics[width=2cm]{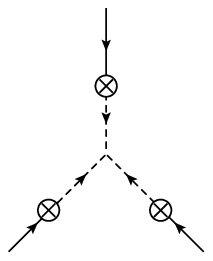}
    \end{minipage}
    \begin{minipage}{0.4\textwidth}
    \centering
    \includegraphics[width=2cm]{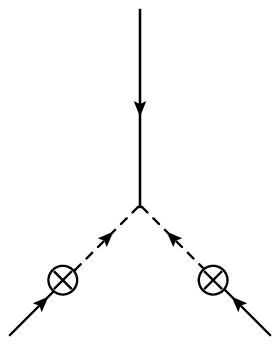}
    \end{minipage}
     \caption{Diagrammatic representations of exchanging non-Gaussianities from entropy mode to non-Gaussianities of adiabatic mode.}
    \label{fig_ng}
    \end{minipage}
    \end{figure}
    This would bring new features to non-Gaussianities such as new
    shapes of momenta configurations \cite{Bartolo:2001cw,cw}.
}



\begin{acknowledgments}
I thank Miao Li, Tao Wang and Yi Wang for useful discussions and
comments. I am grateful to Miao Li for a careful reading of the
manuscript. This work was partly inspired by the project of X. Chen
and Y. Wang. I also would like to thank an anonymous referee for
helping me improve the paper. This work was supported by the NSFC
grant No.10535060/A050207, a NSFC group grant No.10821504 and
Ministry of Science and Technology 973 program under grant
No.2007CB815401.

\end{acknowledgments}



\appendix

\section{``In-in" Formalism}{\label{appsec_inin}}

The ``in-in formalism" (also dubbed as ``Schwinger-Keldysh
formalism", or ``Closed-time path formalism")
\cite{Schwinger:1960qe,Calzetta:1986ey,Jordan:1986ug} (see also
\cite{Weinberg:2005vy,Weinberg:2006ac} for a nice review) is a
perturbative approach for solving the evolution of expectation
values over a finite time interval. It is therefore ideally suited
not only to backgrounds which do not admit an S-matrix description,
such as inflationary backgrounds.

In  the calculation of S-matrix in particle physics, the goal is to
determine the amplitude for a state in the far past $|\psi\ra$ to
become some state $|\psi'\ra$ in the far future,
    \[
        \la \psi' | S |\psi \ra = \la \psi'(+\infty)
        |\psi(-\infty)\ra \,.
    \]
Here, conditions are imposed on the fields at both very early and
very late times. This can be done because that in Minkowski
spacetime, states are assumed to be non-interacting at far past and
at far future, and thus are usually taken to be the free vacuum,
i.e., the vacuum of the free Hamiltonian $H_0$. The free vacuum are
assumed to be in ``one-to-one" correspondence with the true vacuum
of the whole interacting theory, as we adiabatically turn on and
turn off the interactions between $t=-\infty$ and $t=+\infty$.

While the physical situation we are considering here is quite
different. Instead of specifying the asymptotic conditions both in
the far past and far future, we develop a given state \emph{forward}
in time from a specified initial time, which can be chosen as the
beginning of inflation. In the cosmological context, the initial
state is usually chosen as free vacuum, such as Bunch-Davis vacuum,
since at very early times when perturbation modes are deep inside
the Hubble horizon, according to the equivalence principle, the
interaction-picture fields should have the same form as in Minkowski
spacetime.


The Hamiltonian can be split into a free part and an interacting
part: $H=H_0+H_{\textrm{i}}$. The time-evolution operator in the
interacting picture is well-known
    \eq{{\label{U_op}}
        U(\eta_2,\eta_1) = \mathrm{T}\exp\left( -i\int_{\eta_1}^{\eta_2} dt' H_{\textrm{i}\I}(\eta')
        \right)\,,
    }
where subscript ``$\I$" denotes interaction-picture quantities,
$\textrm{T}$ is the time-ordering operator. Our present goal is to
relate the interacting vacuum at arbitrary time $|\Omega_\I(t)\ra$
to the free vacuum $|0_\I\ra$ (e.g., Bunch-Davis vacuum). The trick
is standard. First we may expand $|\Omega_\I(\eta)\ra$ in terms of
eigenstates of free Hamiltonian $H_0$, $|\Omega_\I(\eta)\ra = \sum_n
|n_\I\ra\, \la n_\I| \Omega_\I(\eta)\ra$, then we evolve
$|\Omega_\I(\eta)\ra$ by using (\ref{U_op})
    \eq{{\label{vacuum_evo}}
        |\Omega_\I(\eta_2)\ra = U(\eta_2,\eta_1)
        |\Omega_\I(\eta_1)\ra = |0_\I\ra\,
        \la 0_\I| \Omega_\I\ra + \sum_{n\geq1} e^{+iE_n(\eta_2 - \eta_1)}\, |n_\I \ra\,
        \la n_\I| \Omega_\I(\eta_1)\ra \,.
    }
From (\ref{vacuum_evo}), we immediately see that, if we choose
$\eta_2 = -\infty(1-i\epsilon)$, all excited states in
(\ref{vacuum_evo}) are suppressed. Thus we relate interacting vacuum
at $\eta= -\infty(1-i\epsilon)$ to the free vacuum $|0\ra$ as
    \eq{
        |\Omega_\I(-\infty (1 - i\epsilon) )\ra = |0_\I \ra\,
        \la 0_\I| \Omega_\I \ra
    }
Thus, the interacting vacuum at arbitrary time $\eta$ is given by
    \ea{
        |\textrm{VAC},\textrm{in} \ra &\equiv |\Omega_\I(\eta)\ra =  U(\eta,-\infty(1-i\epsilon))  |\Omega_\I(-\infty(1-i\epsilon))\ra \\
        &= \mathrm{T}\exp\left( -i\int^{\eta}_{-\infty(1-i\epsilon)} d\eta'\, H_{\textrm{i}\I}(\eta')
        \right) |0_\I \ra\,
        \la 0_\I| \Omega_\I \ra \,.
    }


The expectation value of operator $\hat{\mathcal{O}}(\eta)$ at
arbitrary time $\eta$ is evaluated as
    \ea{{\label{in-in_ev}}
        \la \hat{\mathcal{O}}(\eta) \ra &\equiv \frac{ \la \textrm{VAC},\textrm{in} | \hat{\mathcal{O}}(\eta) |\textrm{VAC},\textrm{in}
        \ra }{ \la \textrm{VAC},\textrm{in} |\textrm{VAC},\textrm{in}
        \ra } \\
        &= \soev{ 0_\I }{
        \bar{ \mathrm{T}} \exp\left( i\int^{\eta}_{-\infty{ (1 + i\epsilon)} } d\eta' H_{1\I}(\eta')
        \right) \, \hat{\mathcal{O}}_\I(\eta) \, \mathrm{T}\exp\left( -i\int^{\eta}_{-\infty {( 1-i\epsilon)} } d\eta' H_{\textrm{i}\I}(\eta')
        \right) }{0_\I } \,,
    }
where $\bar{\textrm{T}}$ is the anti-time-ordering operator.

For simplicity, we denote \eq{
    -\infty(1-i\epsilon) \equiv -\infty^+\,,\qquad\qquad -\infty(1+i\epsilon) \equiv
    -\infty^- \,,
    }
 since, e.g., $-\infty^+$
has a positive imaginary part. Now let us focus on the time-order in
(\ref{in-in_ev}). In standard S-matrix calculations, operators
between $\la 0|$ and $|0\ra$ are automatically time-ordered. While
in (\ref{in-in_ev}), from right to left, time starts from infinite
past, or $-\infty^+$ precisely, to some arbitrary time $\eta$ when
the expectation value is evaluated, then back to $-\infty^-$ again.
This time-contour forms a closed-time path, so ``in-in" formalism is
sometimes called ``closed-time path" (CTP) formalism.


The starting point of perturbation theory is the free theory
two-point correlation functions. In canonical quantization
procedure, we write a scalar field as
    \eq{
        \phi_{\vk}(\eta) = u(k,\eta) a_{\vk} + u^{\ast}(k,\eta)
        a^{\dag}_{-\vk} \,,
    }
where $u(k,\eta)$ is the mode function for $\phi_{\vk}(\eta)$ (in
practice, $u_k(\eta)$ and $u^{\ast}_k(\eta)$ are two
linear-independent solutions of equation of motion for
$\phi_{\vk}(\eta)$, which are Wroskian normalized and satisfy some
initial or asymptotic conditions ).

The free two-point function takes the form
    \eq{{\label{2pf}}
        \soev{0}{ \phi_{\vk_1}(\eta_1) \phi_{\vk_2}(\eta_2) }{0}
        \equiv (2\pi)^3 \delta^3(\vk_1+\vk_2)  G_{k_1}(\eta_1,\eta_2)
        \,,
    }
with
    \eq{{\label{free_gf}}
        G_{k_1}(\eta_1,\eta_2) \equiv u_{k_1}(\eta_1)
        u^{\ast}_{k_1}(\eta_2) \,.
    }
In this work, we take (\ref{2pf}) and (\ref{free_gf}) as the
starting point.

Now Taylor expansion of (\ref{in-in_ev}) gives
    \itm{
        \item 0th-order
            \eq{{\label{ev_0th}}
                \lrab{ \hat{\mathcal{O}}(\eta) }^{(0)} = \la 0_\I | \hat{\mathcal{O}}_\I(\eta) |0_\I\ra \,.
                }
        \item 1st-order (one interaction vertex)
            \eq{{\label{app_ev_1st}}
                \lrab{ \hat{\mathcal{O}}(\eta) }^{(1)} = 2\, \textrm{Re} \lrsb{ -i \, \int^{\eta}_{{ -\infty^+} } d\eta'\,
         \soev{0_\I}{ \hat{\mathcal{O}}_\I(\eta)\, H_{\textrm{i}\I}(\eta')  }{0_\I}
            }\,.
            }
         \item 2nd-order (two interaction vertices)
            \eq{{\label{ev_2nd}}
            \begin{aligned}
                \lrab{ \hat{\mathcal{O}}(\eta) }^{(2)} &= - 2\, \textrm{Re} \lrsb{ \int^{\eta}_{{ -\infty^+} } d\eta'\, \int^{\eta'}_{{ -\infty^+}
        } d\eta''\, \soev{0_\I}{ \hat{\mathcal{O}}_\I(\eta)\,  H_{\textrm{i}\I}(\eta')\, H_{\textrm{i}\I}(\eta'') }{0_\I} } \\
        &\qquad\qquad + \int^{\eta}_{{ -\infty^-} } d\eta' \int^{\eta}_{{ -\infty^+} }
        d\eta''\, \soev{0_\I}{ H_{\textrm{i}\I}(\eta') \, \hat{\mathcal{O}}_\I(\eta) \,
         H_{\textrm{i}\I}(\eta'') }{0_\I} \,.
         \end{aligned}
            }
    }

\section{Mathematics}

In this work, we frequently account exponential/sine/cosine-integral
functions, their definitions are
    \ea{
        \text{Ei}(z) &=-\int_{-z}^{\infty } \frac{e^{-t}}{t} \, dt
        \\
        \text{Si}(z) &=\int _0^z \frac{\sin (t)}{t} dt \,,\qquad\qquad \text{Ci}(z)=-\int_z^{\infty } \frac{\cos (t)}{t} \,
        dt \,.
    }
In evaluating various integrals, the following properties are
frequently used: \itm{
    \item For $x>0$,
    \[
        \textrm{Ei}\left(-ix\right)=\text{Ci}(x)-i\left(\text{Si}(x)+\frac{\pi}{2}\right)
        \,,
    \]
    where $\textrm{Ci}(x)$ and $\textrm{Si}(x)$ take real values.
    \item $\textrm{Ei}(-i\infty)=-i\pi$,
    $\textrm{Ei}(+i\infty)=i\pi$, $\textrm{Ci}(+\infty)=0$, $\textrm{Ci}(-\infty)=i\pi$ and
    $\textrm{Si}(\pm \infty)=\pm \frac{\pi}{2}$.
    \item When $x\rightarrow +\infty$
        \eq{
            \textrm{Ei}(-ix) +i\pi = \frac{e^{-i x}}{x} \left( i - \frac{1}{x }+
            \mathcal{O}(x^{-2})\right) \,.
        }
}

\section{Comparison with Previous Results with $c_a = c_e$}{\label{appsec_compare}}

In order to compare calculations in this note to previous results,
we consider a special case $c_a = c_e$, or $\lambda =1$. Our goal is
to show that the field-theoretical perturbative approach taken in
this note is essentially equivalent to previous ``diagonalization"
method taken in \cite{Bartolo:2001vw,Byrnes:2006fr,Lalak:2007vi}. We
take $\Gamma_c$ as example.

It is straightforward calculations to verify that
    \ea{{\label{app_Gamma_c}}
        \Gamma_c(x_{\ast},1) &= \frac{1}{2}\left(x_{\ast}^{2}-1\right)\left[2\text{Ci}(2x_{\ast})\cos(2x_{\ast})-(\pi-2\text{Si}(2x_{\ast}))\sin(2x_{\ast})\right]\\
        &\qquad -x\left[2\text{Ci}(2x_{\ast})\sin(2x_{\ast})+(\pi-2\text{Si}(2x_{\ast}))\cos(2x_{\ast})\right]+2\\
        &\equiv \frac{\pi}{2}x_{\ast}^{3}\left( J_{\frac{3}{2}} \left(x_{\ast}\right) \left.\frac{ \partial J_{\nu}\left(x_{\ast}\right)}{\partial\nu}\right|_{\nu=\frac{3}{2}}+Y_{\frac{3}{2}}\left(x_{\ast}\right) \left.\frac{\partial
        Y_{\nu}\left(x_{\ast}\right)}{\partial\nu}\right|_{\nu=\frac{3}{2}}\right)
        \\
        &=
        \frac{\pi}{4}x_{\ast}^{3}
        \left( \frac{\partial}{\partial\nu} \left|H_{\nu}^{\left(1\right)}\left(x_{\ast}\right)\right|^{2}
        \right)_{\nu=\frac{3}{2}} \,.
        }
Recall that the dimensionless power spectrum for a scalar field in
dS background takes the form
    \eq{
        \mathcal{P}(\eta,k) = \bar{\mathcal{P}} F_{\nu}\left(x\right) \,,
    }
where $\bar{\mathcal{P}} \equiv
\left(\frac{H_{\ast}}{2\pi}\right)^{2}\frac{1}{c_{s}}$ is the
asymptotic super-Hubble limit (i.e. $x_{\ast}\rightarrow 0$) of the
spectrum and $x\equiv -c_s k\eta$ and
    \ea{{\label{app_def_F}}
        F_{\nu}\left(x\right)
        &\equiv\frac{\pi}{2}x^{3}\left|H_{\nu}^{\left(1\right)}\left(x\right)\right|^{2}
        \\
        &=
        \frac{\pi}{2}x^{3}\left|H_{\frac{3}{2}}^{\left(1\right)}\left(x\right)\right|^{2}+\left(\nu-\frac{3}{2}\right)\frac{\pi}{2}x^{3}\left(\frac{\partial}{\partial\nu}\left|H_{\nu}^{\left(1\right)}\left(x\right)\right|^{2}\right)_{\nu=\frac{3}{2}}+\cdots
        \\
        &\equiv  1+x^2  +\left(\nu-\frac{3}{2}\right)f(x) +\cdots
        \,,
    }
with
    \eq{{\label{app_def_f}}
        f(x) \equiv \frac{\pi
        x^{3}}{2\left(1+x^{2}\right)}\left(\frac{\partial}{\partial\nu}\left|H_{\nu}^{\left(1\right)}\left(x\right)\right|^{2}\right)_{\nu=\frac{3}{2}}
        \,.
    }
 Thus in this case the factor $\Gamma_c(x_{\ast},\lambda)$ which
is get from theoretical perturbative calculations is nothing but the
coefficient of the term which is first-order in $\nu-\frac{3}{2}$ in
the expansion of the exact power spectrum:
    \eq{
        \Gamma_c(x_{\ast},1) = \frac{1}{2}f(x_{\ast}) \,.
    }
 Thus
(\ref{app_Gamma_c}) can be compared with (e.g.) Eq.(79) in
\cite{Lalak:2007vi}. This result is as expected, since in this case
the original coupled equations can be diagonalized, and two
approaches (i.e. ``calculating the correlator perturbatively" and
``expanding the exact correlator perturbatively") are essentially
equivalent.

\section{Mode Solutions}{\label{appsec_mode}}

We take the adiabatic equation of motion in (\ref{eom_pert_dec}) as
example. It is convenient to introduce new variable
$x\equiv-c_{a}k\eta$ to characterize the time-evolution{\footnote{A
similar treatment can be found in Appendix A in \cite{Chen:2006nt},
where an alternative variable $y\equiv \frac{c_s k }{aH}$ was chosen
as the evolution variable.}}. After straightforward algebras and
keeping terms up to first-order in slow-varying parameters, the
equation becomes:
    \eq{{\label{app_eom}}
        \frac{d^{2}u_{\sigma}}{dx^{2}}-\frac{s_a}{x}\frac{du_{\sigma}}{dx}
        + \left(1+2s_a-\frac{1}{x^{2}}\left(2+3\epsilon+\frac{3}{2}\eta_{\epsilon}+s_a\right)\right)u_{\sigma}
        = 0\,,
    }
where various slow-varying parameters are defined in
(\ref{sv_para}). Its solution which corresponds to the traditional
BD vacuum is
    \eq{{\label{app_sol}}
        u_{\sigma} = Cx^{\frac{1}{2}\left(1+s_a \right)}H_{\nu_{\sigma}}^{\left(1\right)}\left(\left(1+s_a\right)x\right)
        \,,
    }
where $C$ is the normalization constant to be determined and
$\nu_{\sigma} =
\frac{3}{2}+\frac{1}{2}\left(2\epsilon+\eta_{\epsilon} +
s_a\right)$.

On the other hand, when modes are deep inside the horizon i.e. $x\ll
1$, the above equation becomes
    \eq{
        \frac{d^{2}u_{\sigma}}{dx^{2}}-\frac{s_a}{x}\frac{du_{\sigma}}{dx}+\left(1+2s_a\right)u_{\sigma}=0
        \,,
    }
and its solution which satisfies the Wronskian normalization is
    \eq{
        u_{\sigma}\equiv \left( \bar{H}\left(1-\epsilon\right)\right)^{\frac{s_a}{2}}
        \frac{ e^{i\left(1+s_a\right)x} }{\sqrt{2 (\bar{c}_{s}k)^{1+s_a} }} \,  x^{\frac{s_a}{2}}
        \,,
    }
where we use a ``$\bar{\phantom{A}}$" to denote constant values for
the slow-varying quantities, e.g. $H \sim \bar{H}a^{-\epsilon}$ etc.
Using the fact that for $x\gg 1$, the solution (\ref{app_sol}) has
the following asymptotic expression:
    \eq{
        u_{\sigma} \rightarrow  Cx^{\frac{1}{2}s}e^{i(1+s)x}\sqrt{\frac{2}{\pi}}\frac{e^{-i\frac{\pi}{4}}e^{-\frac{1}{2}i\pi\nu}}{\sqrt{1+s}}
        \,,
    }
the constant in (\ref{app_sol}) can be determined as
    \eq{
        C = \frac{\sqrt{\pi}}{2}\bar{H}^{\frac{s_a}{2}}\left(\bar{c}_{s}k\right)^{-\frac{1+s_a}{2}}\left(1+\frac{s_a}{2}\right)
        e^{\frac{i\pi}{2}\left(\nu_{\sigma}+\frac{1}{2}\right)}
        \,.
    }
This gives the solution in (\ref{free_mode}). It is similar for the
entropy mode solution in (\ref{free_mode}).




\end{document}